\documentclass[pdflatex,sn-basic]{sn-jnl}

\jyear{2021}%

\theoremstyle{thmstyleone}%
\newtheorem{theorem}{Theorem}
\newtheorem{lemma}{Lemma}

\theoremstyle{thmstyletwo}%

\theoremstyle{thmstylethree}%

\usepackage{etoolbox}   
\makeatletter
\patchcmd{\@maketitle}{\artauthors}{{\artauthors}}{}{}
\makeatother
\begin{document}

\title[Withdrawal Success Optimization]{Withdrawal Success Optimization}


\author[1]{\fnm{\quad\quad\quad\quad Hayden} \sur{Brown} ORCID: 0000-0002-2975-2711}\email{haydenb@nevada.unr.edu}

\affil[1]{\orgdiv{Department of Mathematics and Statistics}, \orgname{University of Nevada, Reno}, \orgaddress{1664 \street{N. Virginia Street}, \city{Reno}, \postcode{89557}, \state{Nevada}, \country{USA}}}


\abstract{For $n$ assets and discrete-time rebalancing, the probability to complete a given schedule of investments and withdrawals is maximized over progressively measurable portfolio weight functions. Applications consider two assets, namely the S\&P Composite Index and an inflation-protected bond. The maximum probability and optimal portfolio weight functions are computed for annually rebalanced schedules involving an arbitrary initial investment and then equal annual withdrawals over the remainder of the time period. Applications also consider annually rebalanced schedules that start with dollar cost averaging (equal annual investments) and then shift to equal annual withdrawals. Results indicate noticeable improvements in the probability to complete a given schedule when optimal portfolio weights are used instead of constant portfolio weights like the standard of keeping 90\% in the S\&P Composite Index and 10\% in inflation-protected bonds.}

\keywords{Dollar cost averaging, Terminal wealth, Standard and Poor, L\'evy alpha-stable, Withdrawals}





\maketitle

\section{Introduction}\label{sec1}
Suppose an investor wishes to make a particular schedule of investments and withdrawals. Assume the number of investments and withdrawals are finite and completely determined at time $0$. Given $n$ assets available for investment, the goal is to maximize the probability to complete the schedule, with the maximization occurring over time-adapted portfolio weights. Assume the set of rebalancing times is finite and contains the set of investment and withdrawal times.

Theoretical results present a recursive formula that computes this maximum probability, providing the optimal time-adapted portfolio weights along the way. The recursive formula is also generalized to incorporate a mortality distribution. In this case, the goal is to maximize the probability of completing just the scheduled investments and withdrawals that occur while the investor is alive. 

Applications focus on two assets: the S\&P Composite Index and an inflation protected bond. This combination offers a good balance of high risk - high return and low risk - low return. In particular, the risk and return of the portfolio is increased (decreased) by raising (lowering) the proportion of wealth invested in the S\&P Composite Index. Consequently, the optimal portfolio weights indicate the best time-adapted balance of high risk - high return and low risk - low return for the given schedule of investments and withdrawals.

Two general investment schedules are considered in applications. There is the schedule that starts with a lump sum investment and then has equal withdrawals for a specified length of time or until death. There is also the schedule that starts with dollar cost averaging (DCA) and then has equal withdrawals for a specified length of time or until death. Note that in DCA, equal amounts are invested at equidistant times. In applications, all investments and withdrawals are executed on an annual basis. Rebalancing occurs whenever an investment or withdrawal is made.

The maximum probability to complete the schedule is compared to the probability resulting from portfolio weights that are constant over time. Special attention is given to the $p:(1-p)$ portfolios with $p=.6,\ .9,\ 1$, where $p\%$ of wealth is invested in the S\&P Composite Index and $(1-p)\%$ of wealth is invested in the inflation protected bond. $p=1$ generally gives a probability that is closest to the maximum probability; however, the difference can be several tenths (note that probability is measured here on a scale from 0 to 1). 

The following results from applications demonstrate the advantage in using optimal portfolio weights versus constant portfolio weights. Consider the schedule having an initial lump sum investment and then equal annual withdrawals. Suppose the initial investment is 30 times the withdrawal amount. Then an investor can make 50 annual withdrawals with 95\% confidence, and a 60 year old can make annual withdrawals until death with 99\% confidence. Note that only investing in the S\&P Composite Index would result in 90.9\% and 97.3\% confidence, respectively. Next consider the schedule that starts with DCA and then has equal annual withdrawals. Suppose .5 times the withdrawal amount is invested annually for 30 years. Then the investor can make 50 years of withdrawals with 95\% confidence; only investing in the S\&P Composite Index would result in 92.4\% confidence. Suppose .95 times the withdrawal amount is invested annually for 20 years, starting at age 20. Then starting at age 40, the investor can make annual withdrawals until death, with 95\% confidence; only investing in the S\&P Composite Index would result in 93\% confidence. 

\subsection{Literature Review}
Maximizing the probability that an outcome lies above a particular value is known as the safety first principle, originally considered by \cite{roy1952safety}. Here, the outcome in question is the terminal wealth resulting from a given schedule of investments and withdrawals. Under the setup detailed in section \ref{s:notation}, a given schedule of investments and withdrawals is successfully executed if and only if terminal wealth is non-negative. Thus, the goal is to maximize the probability that terminal wealth is non-negative. In the parlance of \cite{roy1952safety}, the disaster level here is 0. The theory presented here also handles other disaster levels, but the primary focus is on 0. 

To account for nuances of individual choice that safety first misses, \cite{levy2009safety} proposed the expected utility - safety first (EU-SF) model. In particular, the EU - SF model takes a continuous utility function and seeks to maximize the weighted average of that utility and the probability for the outcome to be greater than or equal to some disaster level. Recently, the EU - SF model has been extended to support a probability distortion function, which addresses further nuances of individual choice like the overestimation of low probability events and underestimation of high probability events \citep{li2021portfolio}. Since successful execution of a given schedule of investments and withdrawals is fully characterized by the safety first principle with disaster level 0, there is no need to consider these more complicated extensions here.

The discrete-time portfolio optimization problem is considered in \cite{phelps1962accumulation}, with the goal being to maximize the expected lifetime utility of consumption. The focus is on lifetime utility functions of the form
\begin{equation*}
U(a_1,a_2,...,a_N)=\sum_{i=1}^N\alpha^{i-1}\nu(a_i),\quad 0<\alpha\leq1,
\end{equation*}
where $a_i$ is the amount consumed at the beginning of period $i$ and $\nu$ is bounded, strictly increasing, strictly concave and continuously differentiable. In words, this lifetime utility function is the separable sum of each period's discounted utility of consumption. Expexted lifetime utility of consumption is maximized when there is investment in just one asset that has iid return over each period. The return distribution is assumed to be discrete with a finite number of outcomes. This optimization is generalized in \cite{hakansson1975optimal} to the multi-asset case with bounded, possibly non-discrete, return distributions. Special attention is given the infinite period case and utilities $\nu$ with constant relative or absolute risk aversion index.

The portfolio optimization problem can also be approached from the continuous angle, where stock prices follow stochastic differential equations and portfolios are rebalanced continuously in time. In \cite{karatzas1987optimal}, the goal is to maximize expected discounted utility of consumption and terminal wealth over a continuously rebalanced portfolio. The effect of various wealth-based constraints on the maximization of expected utility of terminal wealth is considered in \cite{basak2001value} and \cite{kraft2013dynamic}. Here, stock price processes can be generated by stochastic differential equations, but the portfolios are not rebalanced continuously in time. It is advantageous to consider discrete-time rebalancing, since there can be significant error in results when a strategy that involves continuous rebalancing is discretized for applications. 

For a portfolio consisting of the S\&P 500 index and US Treasury Bills, \cite{musumeci1999dynamic} use dynamic programming to compute the optimal portfolio weights for an investor looking to maximize his expected utility of wealth $U(W)=-W^{-1}-ae^{-bW}$, where $a$ and $b$ are constants. Results only account for an initial investment that is rebalanced periodically. Additional investments and withdrawals are not considered in the dynamic programming formulation. Here, applications focus on the effect of periodic investments and withdrawals on the optimal portfolio weights, with the goal being to maximize the safety first utility with disaster level 0. Furthermore, this safety first utility is easily understood as the probability to complete a given schedule of investments and withdrawals.

In \cite{bertsekas1996stochastic}, the discrete-time stochastic dynamic programming problem is generalized to the Borel setting where results are carefully treated with measure theory. Theoretical results presented here follow a similar framework. However, the formulation is slightly different. Here, asset prices and portfolio weights are progressively measurable with respect to a filtration that represents the evolution of information over time. 

Results presented here are meant to serve individual investors, ultimately providing an idea of how much retirement income can be reliably generated by an individual account. Maintaining an individual account during retirement can be advantageous because the investor maintains unconditional control over his wealth. Annuities can offer good deals on retirement income, but the purchaser must surrender some control over his wealth. Annuities are also focused on seniors, with many having minimum age requirements that leave out younger investors. Since the results presented here use an individual account, they are more accessible, giving younger investors an annuity-like alternative. For examples of research that considers annuities in the problem of optimizing retirement income, see \cite{antolin2010assessing}, \cite{butt2015effect} and the references therein.

\subsection{Organization}
Section \ref{s:notation} provides the problem set-up. Theoretical results are given in section \ref{s:theory}, with their proofs in Appendix \ref{secA1}. Section \ref{s:app} provides applications of theoretical results using data. Data is described in section \ref{s:data}. Closing remarks and a discussion of related future research ideas are given in section \ref{s:conclusion}.

\section{Preliminaries}\label{s:notation}
Introduce the filtered probability space $(\Omega,\mathcal{F},\mathbb{F},\mathbb{P})$, where $\mathbb{F}:=\{\mathcal{F}(t)\}_{t\in T}$ denotes a filtration of $\mathcal{F}$ and $T\subset[0,\infty)$, $0\in T$. Consider $n$ assets available for investment, each denoted by an index from 1 to $n$. For each $j=1,2,...,n$, let $X_j:T\times\Omega\to(0,\infty)$ be an $\mathbb{F}$-adapted process. When convenient, write $X_j(t)$ in place of $X_j(t,\omega)$, understanding that $X_j(t)$ is an $\mathcal{F}(t)$-measurable function on $\Omega$. In this setting, $X_j(t)$ denotes the value of asset $j$ at time $t$.  

Let $\{t_k\}_{k=0}$ be an increasing sequence in $T$ with $t_0=0$. Require that for each $j$ and $t_k$, $\log X_j(t_{k+1})-\log X_j(t_k)$ is independent of $\mathcal{F}(t_k)$. Suppose $c_k$ is invested or withdrawn at each time step $t_k$. Positive $c_k$ indicate investments, and negative $c_k$ indicate withdrawals. No other investments or withdrawals are made. Further suppose that rebalancing occurs only at those times $t_k$. Note that in this set-up, rebalancing can occur at time $t_k$ with $c_k=0$. 

After accounting for $c_k$, denote the wealth available for investment at time $t_k$ with $W_k$. At each time $t_k$, rebalance $W_k$ according to the $\mathcal{F}(t_k)$-measurable portfolio weight vector $\boldsymbol\pi_k:\Omega\to\Pi$, where $\Pi=\{\mathbf{p}\in[0,1]^n:\sum_{j=1}^np_j=1\}$. When convenient, write $\boldsymbol\pi_k=(\pi_{k1},\pi_{k2},...,\pi_{kn})$, understanding that $\boldsymbol\pi_k$ and each $\pi_{kj}$ is an $\mathcal{F}(t_k)$-measurable function on $\Omega$. In particular, at each time $t_k$, invest $\pi_{kj}W_k$ in asset $j$ for each $j=1,2,...,n$. 

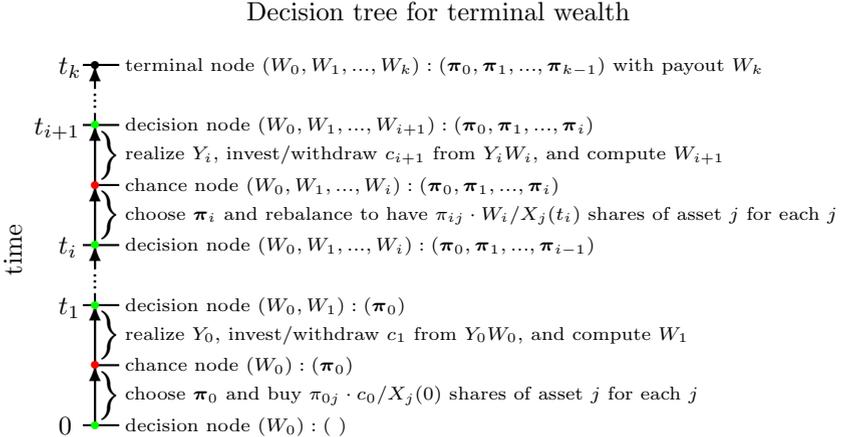
\begin{figure}[h] 
\setlength{\unitlength}{0.4cm}
\begin{picture}(20,15)
\thicklines
\put(9,14.5){Decision tree for terminal wealth}
\put(1,6){\begin{turn}{90}\text{time}\end{turn}}
\put(4,1){\line(0,1){4.4}}
\put(4,5.6){\circle*{.1}}
\put(4,5.8){\circle*{.1}}
\put(4,6.0){\circle*{.1}}
\put(4,6.2){\line(0,1){5.2}}
\put(4,11.6){\circle*{.1}}
\put(4,11.8){\circle*{.1}}
\put(4,12.0){\circle*{.1}}
\put(4,6.2){\vector(0,1){.8}}
\put(4,12.2){\vector(0,1){.8}}
\put(4,1){\line(-1,0){.4}}
\put(4,5){\line(-1,0){.4}}
\put(4,1){\vector(0,1){2}}
\put(4,3){\vector(0,1){2}}
\put(4.8,1){\line(-1,0){.8}}
\put(4.8,3){\line(-1,0){.8}}
\put(4.8,5){\line(-1,0){.8}}
\put(2.8,.7){0}
\put(2.8,4.7){$t_1$}
\put(4,1){\color{green}\circle*{.25}}
\put(4,5){\color{green}\circle*{.25}}
\put(5,.8){{\footnotesize decision node $(W_0):(\ )$}}
\put(5,4.8){{\footnotesize decision node $(W_0,W_1):(\boldsymbol\pi_0)$}}
\put(4.4,1.2){\rotatebox{90}\upbracefill}
\put(4.4,3.2){\rotatebox{90}\upbracefill}
\put(5,1.85){{\footnotesize choose $\boldsymbol\pi_0$ and buy $\pi_{0j}\cdot c_0/X_j(0)$ shares of asset $j$ for each $j$}}
\put(5,3.85){{\footnotesize realize $Y_0$, invest/withdraw $c_1$ from $Y_0W_0$, and compute $W_1$}}
\put(5,2.8){{\footnotesize chance node $(W_0):(\boldsymbol\pi_0)$}}
\put(4,3){\color{red}\circle*{.25}}

\put(4,7){\line(-1,0){.4}}
\put(4,11){\line(-1,0){.4}}
\put(4,7){\vector(0,1){2}}
\put(4,9){\vector(0,1){2}}
\put(4.8,7){\line(-1,0){.8}}
\put(4.8,9){\line(-1,0){.8}}
\put(4.8,11){\line(-1,0){.8}}
\put(2.8,6.7){$t_i$}
\put(2,10.7){$t_{i+1}$}
\put(4,7){\color{green}\circle*{.25}}
\put(4,11){\color{green}\circle*{.25}}
\put(5,6.8){{\footnotesize decision node $(W_0,W_1,...,W_{i}):(\boldsymbol\pi_0,\boldsymbol\pi_1,...,\boldsymbol\pi_{i-1})$}}
\put(5,10.8){{\footnotesize decision node $(W_0,W_1,...,W_{i+1}):(\boldsymbol\pi_0,\boldsymbol\pi_1,...,\boldsymbol\pi_{i})$}}
\put(4.4,7.2){\rotatebox{90}\upbracefill}
\put(4.4,9.2){\rotatebox{90}\upbracefill}
\put(5,7.85){{\footnotesize choose $\boldsymbol\pi_i$ and rebalance to have $\pi_{ij}\cdot W_i/X_j(t_i)$ shares of asset $j$ for each $j$}}
\put(5,9.85){{\footnotesize realize $Y_i$, invest/withdraw $c_{i+1}$ from $Y_iW_i$, and compute $W_{i+1}$}}
\put(5,8.8){{\footnotesize chance node $(W_0,W_1,...,W_{i}):(\boldsymbol\pi_0,\boldsymbol\pi_1,...,\boldsymbol\pi_{i})$}}
\put(4,9){\color{red}\circle*{.25}}

\put(5,12.8){{\footnotesize terminal node $(W_0,W_1,...,W_{k}):(\boldsymbol\pi_0,\boldsymbol\pi_1,...,\boldsymbol\pi_{k-1})$ with payout $W_k$}}
\put(4.8,13){\line(-1,0){.8}}
\put(4,13){\line(-1,0){.4}}
\put(2.8,12.7){$t_k$}
\put(4,13){\circle*{.25}}
\end{picture}
\caption{Illustrates the decision tree used to compute $W_k$. Green points indicate decision nodes, and red points indicate chance nodes.}
\label{fig:dt}
\end{figure}

To simplify notation, let $X_{jk}=X_j(t_{k+1})/X_j(t_{k})$ for $j=1,2,...,n$ and $k=0,1,...$. When convenient, write $\mathbf{X}_k=(X_{1k},X_{2k},...,X_{nk})$. Let $Y_{k}=\sum_{j=1}^n\pi_{kj}X_{jk}$ for $k=0,1,...$. Then the wealth at time step $t_k$ is given by $W_k$, where the $W_k$ are computed recursively via
\begin{equation}
\begin{split}
W_0&=c_0,\\
W_k&=Y_{k-1}W_{k-1}+c_k,\quad k=1,2,...
\end{split}
\label{recursion}
\end{equation}
Again, note that wealth at time step $t_k$ is computed after accounting for the invesment or withdrawal of $c_k$ at time step $t_k$. Furthermore, observe that each $X_{j,k-1}$, $Y_{k-1}$ and $W_k$ is an $\mathcal{F}(t_k)$-measurable function on $\Omega$. Figure \ref{fig:dt} illustrates \eqref{recursion} in the context of a decision tree. 

Some additional conditions must be placed on the $c_k$ in order for $W_k$ to make sense. In particular, $W_k$ will make sense if $W_k(\omega)<0$ implies $W_{k+1}(\omega)<0$ for each $\omega\in\Omega$. This guarantees that a failure to execute a schedule of investments and withdrawals up to time $t_k$, indicated by $W_k(\omega)<0$, will be carried over to time $t_{k+1}$ and indicated by $W_{k+1}(\omega)<0$. To achieve such propagation of negativity in the $W_k$, it suffices to require that $c_k<0$ implies $c_i<0$ for $i>k$. Observe that if $c_0,c_1...,c_k\geq0$, then $W_k(\omega)\geq0$ for all $\omega\in\Omega$. On the other hand, if at least one of the $c_i$ in $c_0,c_1...,c_k$ is negative, then this requirement makes $c_{k+1}$ negative, in which case $W_k(\omega)<0$ implies $W_{k+1}(\omega)<0$ for each $\omega\in\Omega$.

Observe that $W_k$ is a function of each $\boldsymbol\pi_i$ for $i=0,1,...,k-1$. The notation
\begin{equation*}
\underset{\boldsymbol\pi_0,\boldsymbol\pi_1,...,\boldsymbol\pi_{k-1}}{\sup}\mathbb{P}(W_k\geq w)
\end{equation*}
is used to denote the supremum of $W_k$ over all $\mathcal{F}(t_i)$-measurable portfolio weight vectors $\boldsymbol\pi_i$, where $i=0,1,...,k-1$. This kind of abbreviation is used in similar situations where there is a $W_k$-like function that is constructed using the $\mathcal{F}(t_i)$-measurable $\boldsymbol\pi_i$. 

Use $\mathbb{E}[\ \cdot\ ]$ to denote the expectation with respect to $(\Omega,\mathcal{F},\mathbb{P})$. Use $\mathbb{E}[\ \cdot\ \vert\ Z]$ to denote $\mathbb{E}[\ \cdot\ \vert\ \sigma(Z)]$, the expectation conditioned on the $\sigma$-algebra generated by $Z$. Use $\mathbb{R}$ to denote the real numbers. Given $u:\mathbb{R}\to\mathbb{R}$ and $Z:\Omega\to\mathbb{R}$, use $u(Z)$ to denote $u\circ Z$. Given sets $\Psi$, $I$ and $\{(h_i:\Psi\to\mathbb{R}):i\in I\}$, use $(\sup_{i\in I}h_i):\Psi\to\mathbb{R}$ to denote the pointwise supremum of the $h_i$, meaning for each $\psi\in\Psi$, $(\sup_{i\in I}h_i)(\psi)=\sup_{i\in I}(h_i(\psi))$. Let $\mathbf{1}=(1,1,...,1)$ denote the $n$-dimensional vector of 1s, and use $\cdot$ to indicate the dot product. For a vector $\mathbf{a}$, use $\mathbf{a}_j$ to denote the $j$-th component of $\mathbf{a}$. Let $\mathcal{X}_A:\mathbb{R}\to\{0,1\}$ denote the indicator function of $A\subset\mathbb{R}$. 

\begin{lemma}[proposition 2.13 of \cite{de2021martingales}] \label{l:ce}
Let $X:\Omega\to\mathbb{R}^p$ and $Y:\Omega\to\mathbb{R}^q$ be $\mathcal{F}$-measurable functions, $\mathcal{G}$ a sub-$\sigma$-algebra of $\mathcal{F}$ and $h:\mathbb{R}^p\times\mathbb{R}^q\to\mathbb{R}$ a Borel measurable function such that $h(X,Y)$ is integrable. If $X$ is independent from $\mathcal{G}$ and $Y$ is $\mathcal{G}$-measurable, then
\begin{equation*}
\mathbb{E}[h(X,Y)\ \vert\ \mathcal{G}]=H(Y)\quad\text{a.s.},
\end{equation*}
with $H(y)=\mathbb{E}[h(X,y)]$ for all $y\in\mathbb{R}^q$.
\end{lemma}

\begin{lemma}[lemma 1.13 of \cite{KallenbergOlav2019FoMP}] \label{l:dd}
Fix two measurable functions $f$ and $g$ from $\Omega$ into some measurable spaces $(S,\mathcal{S})$ and $(T,\mathcal{T})$, where the former is Borel. Then $f$ is $g$-measurable iff there exists some measurable mapping $h:T\to S$ with $f=h\circ g$.
\end{lemma}

\section{Theoretical Results}\label{s:theory}
For a given non-negative constant $w$ and positive integer $k$, the goal is to find the supremum of $\mathbb{P}(W_k\geq w)$ over the portfolio vectors $\boldsymbol\pi_i$, where $i=0,1,...,k-1$. That goal is expressed in \eqref{eq:argmax}.
\begin{equation}\label{eq:argmax}
\underset{\boldsymbol\pi_0,\boldsymbol\pi_1,...,\boldsymbol\pi_{k-1}}{\sup}\mathbb{P}(W_k\geq w)
\end{equation}
It is unreasonable to compute \eqref{eq:argmax} by testing all possible portfolio vectors $\boldsymbol\pi_i$, where $i=0,1,...,k-1$. Instead, a recursive method is presented that allows more reasonable computation of \eqref{eq:argmax}. 

Before presenting the recursive method used to compute \eqref{eq:argmax}, a simpler method is presented that allows computation of $\mathbb{P}(W_k\geq w)$ for given portfolio vectors $\boldsymbol\pi_i$. This method is useful in applications because it fascilitates comparisons between \eqref{eq:argmax} and the $\mathbb{P}(W_k\geq w)$ resulting from given portfolio vectors $\boldsymbol\pi_i$. 

\subsection{Computing $\mathbb{P}(W_k\geq w)$ for given $\boldsymbol\pi_i$}\label{s:givenpi}
The simpler recursive method arises after expressing $\mathbb{P}(W_k\geq w)$ in terms of conditional expectations. Define the Borel measurable function $u_k:\mathbb{R}\to\mathbb{R}$ such that 
\begin{equation}\label{eq:uk}
u_k(x)=\begin{cases}1,&x\geq w\\
0,&\text{otherwise}
\end{cases}
\end{equation}
Using lemma \ref{l:dd}, let $u_i:\mathbb{R}\to\mathbb{R}$, $i=0,1,...,k-1$, denote the Borel measurable functions satisfying
\begin{equation}\label{eq:ui}
u_i(W_i)=\mathbb{E}[u_{i+1}(W_{i+1})\ \vert\ W_i]\quad\text{a.s.}
\end{equation}
Note that integrability of $u_{i+1}(W_{i+1})$ follows from induction and the fact that $u_k$ is bounded. Furthermore, $W_{i+1}$ is $\mathcal{F}$-measurable and $\sigma(W_i)$ is a sub $\sigma$-algebra of $\mathcal{F}$, so the conditional expectation in \eqref{eq:ui} is a well-defined $\sigma(W_i)$-measurable function. Next observe that
\begin{equation}\label{eq:etoe}
\mathbb{P}(W_k\geq w)=\mathbb{E}[u_k(W_k)]. 
\end{equation}
By \eqref{eq:ui} and the law of total expectation,
\begin{equation}\label{eq:uce}
\mathbb{E}[u_k(W_k)]=\mathbb{E}[u_{k-1}(W_{k-1})]=\ .\ .\ .\ =\mathbb{E}[u_0(W_0)].
\end{equation}
Since $W_0=c_0$ is deterministic, $\mathbb{E}[u_0(W_0)]=u_0(c_0)$. It follows from \eqref{eq:etoe} and \eqref{eq:uce} that 
\begin{equation}\label{eq:ptou}
\mathbb{P}(W_k\geq w)=u_0(c_0).
\end{equation}
If the $\boldsymbol\pi_i$ are $\sigma(W_i)$-measurable for $i=0,1,...,k-1$, then by lemma \ref{l:dd}, there exist Borel measurable $\boldsymbol\phi_i:\mathbb{R}\to\Pi$ such that $\boldsymbol\pi_i=\boldsymbol\phi_i(W_i)$ for each $i$. In this situation, $u_0(c_0)$ can be computed recursively using \eqref{eq:ui} and the following. Recall from \eqref{recursion} that $W_{i+1}=(\boldsymbol\pi_i\cdot\mathbf{X}_i)W_i+c_{i+1}$. If $\boldsymbol\pi_i=\boldsymbol\phi_i(W_i)$ for each $i$, it follows from lemma \ref{l:ce} that there is an $S\in\mathcal{F}$ with $\mathbb{P}(S)=0$ such that for each $x\in \{W_i(\omega):\omega\in\Omega\setminus S\}$,
\begin{equation}\label{eq:ui2}
u_i(x)=\mathbb{E}[u_{i+1}((\boldsymbol\phi_i(x)\cdot\mathbf{X}_i)x+c_{i+1})].
\end{equation} 
For simplicity, compute $u_i(x)$ using $\eqref{eq:ui2}$ for all $x\in\mathbb{R}$. Next, the above recursion is modified to incorporate the supremum in \eqref{eq:argmax}. 

\subsection{Computing \eqref{eq:argmax}}
First define the non-decreasing upper semicontinuous function $v_k:\mathbb{R}\to[0,1]$ such that 
\begin{equation}\label{eq:vk}
v_k(x)=\begin{cases}1,&x\geq w\\
0,&\text{otherwise}
\end{cases}
\end{equation}
Let $v_i:\mathbb{R}\to[0,1]$, $i=0,1,...,k-1$, denote the non-decreasing upper semicontinuous functions satisfying
\begin{equation}\label{eq:VI}
v_i(x)=\max_{\mathbf{r}\in\Pi}\mathbb{E}[v_{i+1}((\mathbf{r}\cdot\mathbf{X}_i)x+c_{i+1})]\quad\forall x\in\mathbb{R}.
\end{equation}
Moreover, \eqref{eq:argmax} is given by $v_0(c_0)$, which can be computed recursively, starting with \eqref{eq:vk} and then using \eqref{eq:VI}. Proofs of the previous statements are given in sections \ref{s:Av} and \ref{s:Av0} of the appendix. 

\subsection{Computing \eqref{eq:argmax} with stock-bond portfolios}\label{s:sb}
Fix $T=[0,\infty)$ and $n=2$. Let $X_1(t)$ denote the value of the stock at time $t$. Assume the $X_{1i}$ are continuous in the sense that $\mathbb{P}(X_{1i}=x)=0$ for each $x>0$ and $i=0,1,...,k-1$. Let $X_2(t)=(1+r)^t$, meaning $X_2(t)$ denotes the value of the bond, with interest $r\geq0$, at time $t$. Let $\mathbb{F}$ be the natural filtration generated by $(X_1(t),X_2(t))$. Let $w_k=w$, and for $i=0,1,...,k-1$, let 
\begin{equation}\label{eq:wir}
w_i=\frac{w_{i+1}-c_{i+1}}{1+r}.
\end{equation}
If $w_i\leq0$, then only investing in the bond would yield $\mathbb{P}(W_k\geq w)=1$. So the case where $w_i\leq0$ is not worth studying. Hence, require that each $w_i$ ($i=0,1,...,k-1$) is positive. 

Under the above set-up, each $v_i$ ($i=0,1,...,k-1$) is continuous over $\mathbb{R}\setminus\{w_i\}$ and right continuous at $w_i$. When $x\geq w_i$,  
\begin{equation*}
v_i(x)=\mathbb{E}[v_{i+1}(((0,1)\cdot\mathbf{X}_i)x+c_{i+1})]=1.
\end{equation*}
For $x\geq0$,
\begin{equation}\label{eq:visb}
v_i(x)=\max_{\mathbf{r}\in\Pi}\int_{A_{\mathbf{r}}}v_{i+1}((\mathbf{r}\cdot\mathbf{X}_i)x+c_{i+1})d\mathbb{P},
\end{equation}
where $A_{\mathbf{r}}=\{\omega:0\leq(\mathbf{r}\cdot\mathbf{X}_i)x+c_{i+1}\}$. See section \ref{s:Asb} in the appendex for proofs of the previous three statements. 

It follows from \eqref{eq:vk} and \eqref{eq:visb} that for $x\in(0,w_{k-1})$,
\begin{equation}\label{eq:vkm1}
\begin{split}
v_{k-1}(x)&=\max_{\mathbf{r}\in\Pi}\mathbb{P}((\mathbf{r}\cdot\mathbf{X}_{1,k-1})x+c_k\geq w)\\
&=\max_{q\in(0,1]}\mathbb{P}\Big(\frac{X_{1,k-1}}{1+r}\geq 1+\frac{1}{q}\big(\frac{w_{k-1}}{x}-1\big)\Big)\\
&=\mathbb{P}\Big(\frac{X_{1,k-1}}{1+r}\geq \frac{w_{k-1}}{x}\Big).
\end{split}
\end{equation}

\subsection{Extension to a mortality distribution}
Let the Borel measurable function $\tau:\Omega\to(0,\infty)$ denote the investor's time of death. Require that $\tau$ is independent of $X_j(t)$ for every $j=1,2,...,n$ and $t\in T$. Assume $\tau$ is a continuous random variable so that borderline cases where death occurs at time $t_i$ is not an issue. Let $\tau_d:\Omega\to\{0,1,...\}$ be such that $\tau(\omega)\in(t_i,t_{i+1}]$ implies $\tau_d(\omega)=i$. By the law of total probability,
\begin{equation}\label{eq:utau}
\mathbb{P}(W_{\tau_d}\geq w)=\sum_{i=0}^{\infty}\mathbb{P}(W_i\geq w,\ \tau\in(t_i,t_{i+1}]).
\end{equation}
If the $\boldsymbol\pi_i$ are $\sigma(W_i)$-measurable, then $\tau$ is independent to each $W_i$ and
\begin{equation}\label{eq:utauIDP}
\mathbb{P}(W_{\tau_d}\geq w)=\sum_{i=0}^{\infty}\mathbb{P}(W_i\geq w)\mathbb{P}(\tau\in(t_i,t_{i+1}]).
\end{equation}
Moreover, each $\mathbb{P}(W_i\geq w)$ in \eqref{eq:utauIDP} can be computed using section \ref{s:givenpi} or direct simulation of the $W_i$.

Now it is tempting to maximize \eqref{eq:utau} over $\boldsymbol\pi_0,\boldsymbol\pi_1,...$. First set up the recursion
\begin{equation}
\begin{split}
\overline{W}_0&=c_0,\\
\overline{W}_i&=(1-\mathcal{X}_{(0,t_i]}(\tau))(Y_{i-1}\overline{W}_{i-1}+c_i)\\
&\quad+\mathcal{X}_{(0,t_i]}(\tau)\overline{W}_{i-1},\quad i=1,2,...,k.
\end{split}
\label{Wbar}
\end{equation}
Observe that
\begin{equation}\label{eq:Wbarcases}
\overline{W}_i(\omega)=\begin{cases}W_i(\omega)&\tau(\omega)>t_i,\\
\overline{W}_{i-1}(\omega)&\text{otherwise},
\end{cases}
\end{equation}
which implies
\begin{equation}\label{eq:wkbar}
\mathbb{P}(\overline{W}_k\geq w)=\mathbb{P}(W_k\geq w,\ \tau>t_k)+\sum_{i=0}^{k-1}\mathbb{P}(W_i\geq w,\ \tau\in(t_i,t_{i+1}]).
\end{equation}
If $\mathbb{P}(\tau>t_k)=0$, then \eqref{eq:wkbar} and \eqref{eq:utau} coincide. Otherwise, $k$ can be chosen large enough such that \eqref{eq:wkbar} approximates \eqref{eq:utau}. So for $k$ sufficiently large,
\begin{equation}\label{eq:maxWbar}
\underset{\boldsymbol\pi_0,\boldsymbol\pi_1,...}{\sup}\mathbb{P}(W_{\tau_d}\geq w)\approx\underset{\boldsymbol\pi_0,\boldsymbol\pi_1,...,\boldsymbol\pi_{k-1}}{\sup}\mathbb{P}(\overline{W}_k\geq w).
\end{equation}
There is also the inequality
\begin{equation}\label{eq:Wbarlb}
\underset{\boldsymbol\pi_0,\boldsymbol\pi_1,...}{\sup}\mathbb{P}(W_{\tau_d}\geq w)\geq\underset{\boldsymbol\pi_0,\boldsymbol\pi_1,...,\boldsymbol\pi_{k-1}}{\sup}\mathbb{P}(\overline{W}_k\geq w)-\mathbb{P}(\tau>t_k),
\end{equation}
which holds for any $k$. Notice the similarity between \eqref{eq:argmax} and the right side of \eqref{eq:maxWbar}. This similarity is taken advantage of to obtain the following result.

Let $\tau_i:\Omega\to\{0,1\}$ be independent $\text{Bernoulli}(p_i)$ random variables for $i=0,1,...$, where $p_i\in[0,1]$. Additionally require that each $\tau_i$ is $\mathcal{F}(t_{i+1})$-measurable and independent of $X_j(t)$ for every $j=1,2,...,n$ and $t\in T$. Define $\tau$ such that for each $\omega\in\Omega$ and $i=0,1,...$,
\begin{equation*}
\tau(\omega)\in(t_i,t_{i+1}]\iff\sum_{j=0}^i\tau_j(\omega)=1. 
\end{equation*}
Let $\overline{v}_k=v_k$, where $v_k$ is as in \eqref{eq:vk}. For $i=0,1,...,k-1$, let $\overline{v}_i:\mathbb{R}\to[0,1]$ denote the non-decreasing upper semicontinuous functions satisfying
\begin{equation}
\begin{split}
\overline{v}_i(x)&=(1-p_i)\max_{\mathbf{r}\in\Pi}\mathbb{E}[\overline{v}_{i+1}((\mathbf{r}\cdot\mathbf{X}_i)x+c_{i+1})]+p_iv_k(x),\quad\forall x\in\mathbb{R}.
\end{split}
\label{eq:evbarix}
\end{equation}
Then the right side of \eqref{eq:maxWbar} is given by $\overline{v}_0(c_0)$, which can be computed recursively, starting with \eqref{eq:vk} (since $\overline{v}_k=v_k$) and using \eqref{eq:evbarix} after that. Proof of the previous statements are given in section \ref{s:Avbar} of the appendix. 

The following details some useful facts about the $\overline{v}_i$.
\begin{itemize}
\item The recursion used to obtain $\overline{v}_0(c_0)$ can be kick-started at $k-1$ by use of $v_{k-1}$. Taking $i$ to be $k-1$ in \eqref{eq:evbarix} yields
\begin{equation}\label{eq:evbarkm1}
\overline{v}_{k-1}(x)=(1-p_{k-1})v_{k-1}(x)+p_{k-1}v_k(x)\quad\forall x\in\mathbb{R}.
\end{equation}
\item Observe that if $\mathbb{P}(\tau\leq t_{i+1})=0$, then \eqref{eq:evbarix} reduces to 
\begin{equation*}
\overline{v}_i(x)=\max_{\mathbf{r}\in\Pi}\mathbb{E}[\overline{v}_{i+1}((\mathbf{r}\cdot\mathbf{X}_i)x+c_{i+1})]\quad\forall x\in\mathbb{R}.
\end{equation*}
\item Under the set-up of section \ref{s:sb}, $\overline{v}_i(x)=1$ for $x\geq w_i$, and for $x\geq0$ there is
\begin{equation*}
\mathbb{E}[\overline{v}_{i+1}((\mathbf{r}\cdot\mathbf{X}_i)x+c_{i+1})]=\int_{A_{\mathbf{r}}}\overline{v}_{i+1}((\mathbf{r}\cdot\mathbf{X}_i)x+c_{i+1})d\mathbb{P}.
\end{equation*}
\end{itemize}

\section{Applications}\label{s:app}
Theoretical results are applied to portfolios consisting of the S\&P Composite Index and an inflation-protected bond. Applications use a mortality distribution that is based on the 2017 per-age death rates of the US Social Security area population. Section \ref{s:data} describes the S\&P Composite Index data and the mortality distribution data. Section \ref{s:appSetup} justifies the treatment of S\&P returns as iid and details the set-up needed to apply theoretical results. Algorithms used in applications are described in section \ref{s:appSetup}. Results of applications are given in section \ref{s:resultsA}.

\subsection{Data}\label{s:data}
Annual data from the S\&P Composite Index and Comsumer Price Index is taken from \url{http://www.econ.yale.edu/~shiller/data.htm}, collected for easy access at \url{https://github.com/HaydenBrown/Investing}. The data spans 1871 to 2020 and is described in table \ref{t:data}. Note that S\&P Composite Index refers to Cowles and Associates from 1871 to 1926, Standard \& Poor 90 from 1926 to 1957 and Standard \& Poor 500 from 1957 to 2020. Cowles and Associates and the S\&P 90 are used here as backward extensions of the S\&P 500.

The data is transformed so that annual returns incorporate dividends and are adjusted for inflation. In particular, returns are computed using the consumer price index, the S\&P Composite Index price and the S\&P Composite Index dividend. Use the subscript $k$ to denote the $k$th year of $C$, $I$ and $D$ from Table \ref{t:data}. The return for year $k$ is computed as $\frac{I_{k+1}+D_k}{I_k}\cdot\frac{C_k}{C_{k+1}}$. 

\begin{table}
\begin{center}
\caption{Data variable descriptions.}\label{t:data}
\begin{tabular}{ ll } 
\toprule
Notation & Description \\
\toprule
 I & Average monthly close of the S\&P composite index \\ 
 D & Dividend per share of the S\&P composite index \\ 
 C & January consumer price index \\ 
\toprule
\end{tabular}
\end{center}
\end{table}

\begin{figure}
  \includegraphics[width=\linewidth]{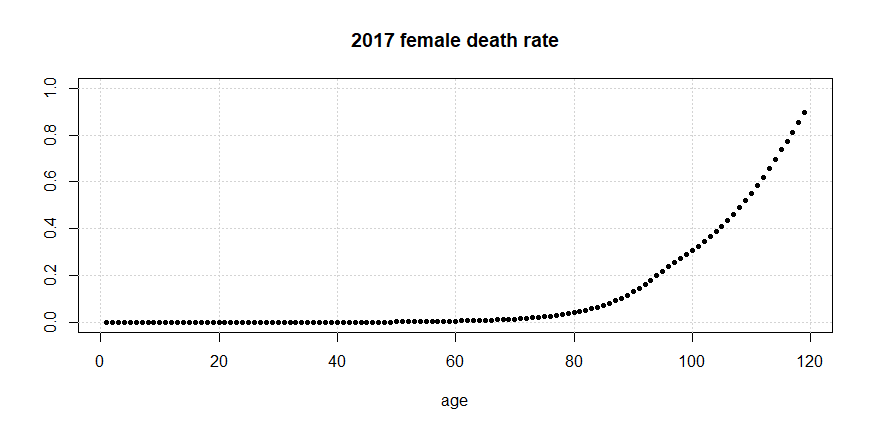}
  \caption{2017 female death rates for the US Social Security area population.}
  \label{fig:fdr}
\end{figure}

Death rates are taken from \url{https://www.ssa.gov}, the official website of the Social Security Administration. In particular, the female per-age death rates of the US Social Security area population are taken from the 2017 period life table. Female death rates are used because they are generally lower than male death rates. The female death rates are illustrated in figure \ref{fig:fdr}.
Let $d_j$ denote the 2017 female death rate for age $j$, and let $s$ denote the starting age for a given schedule of investments and withdrawals. Applications use $k=120-s$ and
\begin{equation*}
p_i=d_{s+i},\quad i=0,1,...,k-1.
\end{equation*}

\subsection{Set-up}\label{s:appSetup}
In order to apply theoretical results, $T$, $n$, $X_j(t)$ for $j=1,2,...,n$, $\mathbb{F}$ and $\{t_k\}_{k=0}$ need to be specified. Set $T=[0,\infty)$ and $n=2$. Let $X_1(t)$ denote the inflation-adjusted value of the S\&P Composite Index at time $t$, and let $X_2(t)=(1+r)^t$, meaning $X_2(t)$ denotes the inflation-adjusted value of an inflation-protected bond, with interest $r$, at time $t$. Set $t_k=k$ for $k=0,1,...$. When no mortality distribution is present, let $\mathbb{F}$ be the natural filtration generated by $(X_1(t),X_2(t))$. When a mortality distribution is present, adjust the previous $\mathbb{F}$ as follows
\begin{equation*}
\mathcal{F}(t)\gets\sigma\Big(\mathcal{F}(t)\cup\bigcup_{i-1\leq t}\sigma(\tau_i)\Big),\quad t\in T.
\end{equation*}

Since $X_1(t)$ and $X_2(t)$ are inflation-adjusted, it follows that the $c_k$ and $W_k$ are also inflation-adjusted. For example, if $c_k=-2$ and inflation is 5\% from time 0 to time $t_k$, then the actual amount withdrawn at time $t_k$ is $2\cdot 1.05$. In other words, 2 is the inflation-adjusted amount withdrawn, and $2\cdot 1.05$ is the actual amount withdrawn. 

\begin{figure}
  \includegraphics[width=\linewidth]{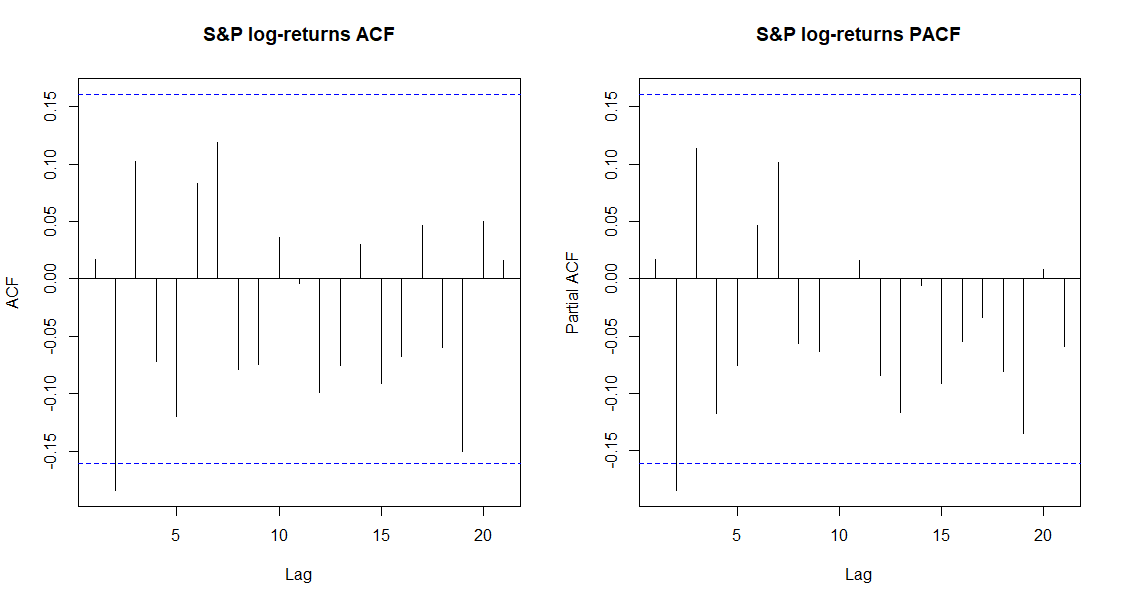}
  \caption{Left: The autocorrelation function of annual S\&P log-returns. Right: The partial autocorrelation function of annual S\&P log-returns.}
  \label{fig:acf}
\end{figure}

Theoretical results also require the $\mathbf{X}_k$ to be independent of $\mathcal{F}(t_k)$. It suffices to have iid $\mathbf{X}_k$. Observing that $X_2(t)$ is deterministic, it suffices to justify the assumption that annual S\&P returns are iid. Table \ref{t:ljung} provides p-values from the Ljung-Box test on returns, log-returns and the absolute value of annual log-returns. The p-value of .078 is slightly concerning, but not enough to reject independence with the usual 95\% confidence. The autocorrelation and partial autocorrelation functions of annual log-returns are given in figure \ref{fig:acf}.  Overall, the assumption that annual S\&P returns are independent appears to be supported by the data. Next, the distribution of the annual returns is fitted.

\begin{table}
\begin{center}
\caption{p-values of Ljung-Box test on annual S\&P returns}\label{t:ljung}
\begin{tabular}{ lccc } 
\toprule
 & returns & log-returns & absolute log-returns\\
\toprule
 lag = 1 & .928 & .833 & .555\\ 
 lag = 5 & .113 &.078 & .975\\ 
\toprule
\end{tabular}
\end{center}
\end{table}

\begin{figure}
  \includegraphics[width=\linewidth]{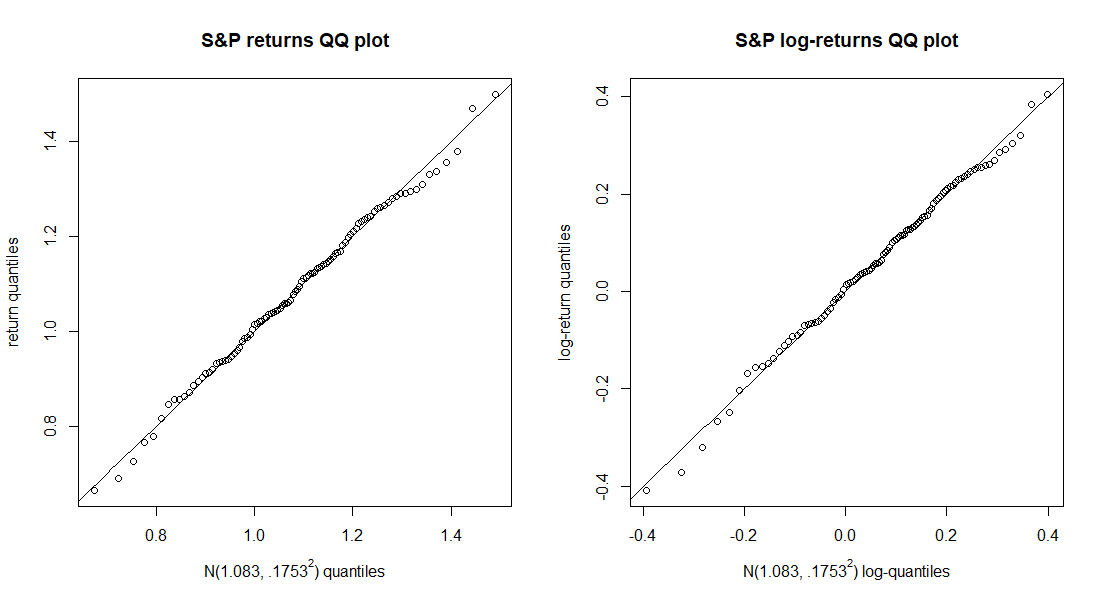}
  \caption{Left: Quantiles of annual S\&P returns versus $\mathcal{N}(1.083,.1753^2)$ quantiles. Right: Quantiles of annual S\&P log-returns versus $\mathcal{N}(1.083,.1753^2)$ log-quantiles. The line indicates where points should be if quantiles are identical.}
  \label{fig:qq}
\end{figure}

\begin{figure}
  \includegraphics[width=\linewidth]{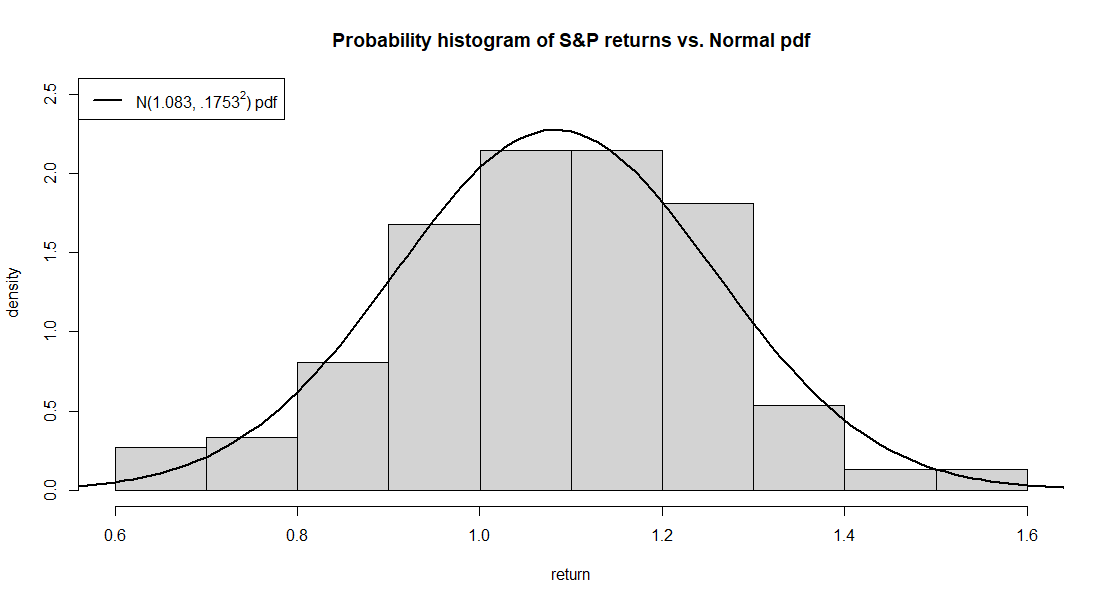}
  \caption{Probability historgram for the annual S\&P return data. The probability density function of $\mathcal{N}(1.083,.1753^2)$ is superimposed for comparison.}
  \label{fig:pdf}
\end{figure}

A simple and effective option is to fit returns to a Normal distribution with mean and standard deviation equal to their corresponding sample values (see figures \ref{fig:qq} and \ref{fig:pdf}). This fit works especially well because the mean and standard deviation of its logarithm are approximately equal to their corresponding sample values as well. In particular, if $X_{1k}\sim N(1.083,.1753^2)$, then the mean and standard deviation of $\log X_{1k}$ are approximately equal to the sample mean (.06578) and standard deviation (.1690) of annual S\&P log-returns. The mean (.06534) and standard deviation (.1680) of $\log X_{1k}$ were computed as the sample mean and standard deviation of 100,000 samples of $\log X_{1k}$. Note that under this fitted distribution, it is possible to realize a negative return. However, the probability of realizing a negative return is so small (on the order of $10^{-10}$), that it is negligible here. 

Now fix $k$ and require that $w_i>0$ for $i=0,1,...,k-1$, where $w_i$ is as in section \ref{s:sb}.

\subsubsection{Simulating $\mathbb{P}(W_k\geq w)$ and $\mathbb{P}(\overline{W}_k\geq w)$} \label{s:usim}
Algorithms \ref{alg:usim} and \ref{alg:uhatsim} compute $\mathbb{P}(W_k\geq w)$ and $\mathbb{P}(\overline{W}_k\geq w)$, respectively, via simulation. Both algorithms use $\boldsymbol\pi_{i}$ that are defined using Borel measurable $q_i:\mathbb{R}\to[0,1]$. In particular, algorithm \ref{alg:usim} uses
\begin{equation*}
\boldsymbol\pi_{i}=(q_i(W_i),1-q_i(W_i)),\quad i=0,1,...,k-1,
\end{equation*}
and algorithm \ref{alg:uhatsim} uses
\begin{equation*}
\boldsymbol\pi_{i}=(q_i(\overline{W}_i),1-q_i(\overline{W}_i)),\quad i=0,1,...,k-1.
\end{equation*}
In algorithm \ref{alg:usim}, $N$ realizations of $W_k$ are simulated, and then $\mathbb{P}(W_k\geq w)$ is computed as the number of realizations greater than or equal to $w$, divided by $N$. Likewise, in algorithm \ref{alg:uhatsim}, $N$ realizations of $\overline{W}_k$ are simulated, and then $\mathbb{P}(\overline{W}_k\geq w)$ is computed as the number of realizations greater than or equal to $w$, divided by $N$.  

\subsubsection{Computing $v_0(c_0)$}\label{s:valg} 
Let $M$ be a sufficiently large postive integer. In algorithm \ref{alg:v}, $v_i(x)$ is computed recursively for $x\in D_i$, where
\begin{equation*}
D_i=\Big\{\frac{mw_i}{M}:m=1,...,2M\Big\}.
\end{equation*}
The following elaborates on the details behind algorithm \ref{alg:v}. 

Recall that $v_k$ is given by \eqref{eq:vk}. Observe that the set-up detailed at the top of section \ref{s:appSetup} aligns with that of section \ref{s:sb}. So $v_i(x)=1$ for $x\geq w_i$ and $i=0,1,...,k$. Furthermore, $v_{k-1}(x)$ is given by \eqref{eq:vkm1} for $x\in(0,w_{k-1})$.

Denote the pdf and cdf of the iid $X_{1i}$ with $f$ and $F$, respectively. Then \eqref{eq:visb} implies that for each $x\in(0,w_i)$, $v_i(x)$ is the maximum of $v_{i+1}((1+r)x+c_{i+1})$ and 
\begin{equation}\label{eq:vqf}
\max_{q\in(0,1]}\int_a^{\infty}v_{i+1}((qz+(1-q)(1+r))x+c_{i+1})f(z)dz,
\end{equation}
where $q$ indicates the proportion invested in the stock at time $t_i$ and
\begin{equation*}
a=1+r-\frac{1+r}{q}-\frac{c_{i+1}}{qx}.
\end{equation*}
Transforming the integral in \eqref{eq:vqf} with the substitution 
\begin{equation*}
y=(qz+(1-q)(1+r))x+c_{i+1}
\end{equation*} 
yields
\begin{equation}\label{eq:vqfT}
\frac{1}{qx}\int_0^\infty v_{i+1}(y)f\Big(1+r-\frac{1+r}{q}+\frac{y-c_{i+1}}{qx}\Big)dy.
\end{equation}
Summarizing,
\begin{equation}\label{eq:visbF}
v_i(x)=\max\big\{v_{i+1}((1+r)x+c_{i+1}),\ \max_{q\in(0,1]}\eqref{eq:vqfT}\big\}.
\end{equation}

Let $\theta=(1+r)x+c_{i+1}$. Algorithm \ref{alg:v} first computes $v_{i+1}(\theta)$, which is needed in \eqref{eq:visbF}. If $\theta\leq0$, then $c_{i+1}<0$, and it follows that $v_{i+1}(\theta)=0$. If $0\leq\theta\leq\min_{y\in D_{i+1}}y$, then a lower bound of $v_{i+1}(\theta)$ is given by $0$. If $\min_{y\in D_{i+1}}y\leq\theta<w_{i+1}$, then a lower bound of $v_{i+1}(\theta)$ is given by $v_{i+1}(y^*)$, where $y^*$ is the closest lesser element in $D_{i+1}$ to $\theta$. If $w_{i+1}\leq \theta$, then $v_{i+1}(\theta)=1$.

Algorithm \ref{alg:v} computes \eqref{eq:vqfT} as
\begin{equation}\label{eq:vqfTS}
(1-F(a))\cdot\frac{\sum_{y\in D_{i+1}}v_{i+1}(y)f\Big(1+r-\frac{1+r}{q}+\frac{y-c_{i+1}}{qx}\Big)}{\sum_{y\in D_{i+1}}f\Big(1+r-\frac{1+r}{q}+\frac{y-c_{i+1}}{qx}\Big)}.
\end{equation}
Note that the fraction in \eqref{eq:vqfTS} approximates the expectation of $v_{i+1}(Y)$ given $Y>0$, where $Y$ has pdf
\begin{equation*}
\frac{1}{qx}\cdot f\Big(1+r-\frac{1+r}{q}+\frac{y-c_{i+1}}{qx}\Big).
\end{equation*}
Furthermore, $\mathbb{P}(Y>0)=1-F(a)$.

The maximum over $q$ in \eqref{eq:visbF} is computed with an iterated grid search, where the grid is refined at each iteration. In particular, the first grid tests $q$ in $G_1=\{.01,.02,...,.99\}$. Let $q_1$ denote the $q$ in $G_1$ that produces the maximum of $\eqref{eq:vqfTS}$. The next grid is $G_2=\{q_1\pm .001m: m=0,1,...,9\}$. Let $q_2$ denote the $q$ in $G_2$ that produces the maximum of $\eqref{eq:vqfTS}$. The last grid is $G_3=\{q_2+.0001m:m=-9,-8,...,10\}$. Let $q_3$ denote the $q$ in $G_2$ that produces the maximum of $\eqref{eq:vqfTS}$. From here, algorithm \ref{alg:v} uses the approximation
\begin{equation*}
\max_{q\in(0,1]}\eqref{eq:vqfT}\approx\eqref{eq:vqfTS}\vert_{q=q_3}.
\end{equation*}

\subsubsection{Computing $\widehat{v}_0(c_0)$}
Algorithm \ref{alg:vhat} computes $\overline{v}_0(c_0)$ in a similar fashion to how algorithm \ref{alg:v} computes $v_0(c_0)$. The differences are outlined below. 
\begin{itemize}
\item $\overline{v}_{k-1}(x)$ is given by 
\begin{equation*}
(1-p_{k-1})\cdot\eqref{eq:vkm1}+p_{k-1}v_k(x)
\end{equation*}
for $x\in (0,w_i)$. 
\item $\overline{v}_i$ is computed using \eqref{eq:evbarix} for $i=0,1,...,k-2$. The expectation in \eqref{eq:evbarix} is computed via \eqref{eq:vqfTS} (after replacing $v$ with $\overline{v}$). Like in algorithm \ref{alg:v}, the maximum in \eqref{eq:evbarix} is computed by comparing $\mathbf{r}=(0,1)$ with the optimal $\mathbf{r}=(q,(1-q))$ coming from the iterated grid search. 
\item The statements about $v_{i+1}(\theta)$ given in section \ref{s:valg} also hold after replacing $v$ with $\overline{v}$.
\end{itemize}

\subsubsection{Simulating \eqref{eq:argmax} and the right side of \eqref{eq:maxWbar}}\label{s:lininterp}
First consider simulation of \eqref{eq:argmax}. Execute algorithm \ref{alg:v} and return the $q^*_i(x)$ for $x\in D_i$. The values of $q^*_i(x)$ for $x\notin D_i$ are computed via the following linear interpolation. Set 
\begin{equation*}
q^*_i(x)=\begin{cases}
1&x\leq0\\
0&x\geq w_i.
\end{cases}
\end{equation*}
For $x\in (0,w_i)\setminus D_i$, let $y_x$ denote the largest element of $D_i\cup\{0\}$ that is less than or equal to $x$, and set
\begin{equation*}
q^*_i(x)=q^*_i(y_x)+\frac{x-y_x}{w_i/M}\cdot(q^*_i(y_x+w_i/M)-q^*_i(y_x)).
\end{equation*}
To simulate \eqref{eq:argmax} simply use the $q^*_i$ described above as the $q_i$ in algorithm \ref{alg:usim}.

The procedure to simulate the right side of \eqref{eq:maxWbar} is very similar. Follow the above procedure after replacing algorithms \ref{alg:v} and \ref{alg:usim} with \ref{alg:vhat} and \ref{alg:uhatsim}, respectively.

\subsection{Results}\label{s:resultsA}
First consider the case where a mortality distribution is not present, and a constant annual withdrawal of 1 unit is made for $k$ years, after an initial investment ($c_0>0$ and $c_i=-1$ for $i=1,2,...,k$). 

\begin{figure}
  \includegraphics[width=\linewidth]{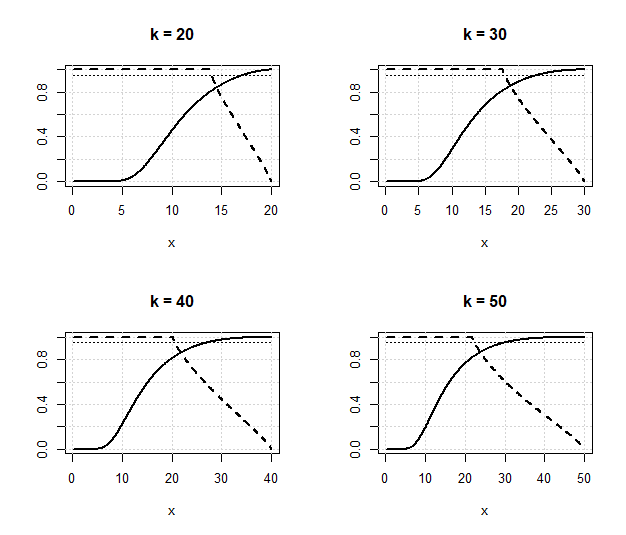}
  \caption{Using algorithm \ref{alg:v} with $M=300$ and $w=r=0$, illustrates $v_0(x)$ (thick line) and $q_i^*(x)$ (thick dashed line) for various $k$, provided $X_{1i}\sim\mathcal{N}(1.083,.1753^2)$ and $c_{i+1}=-1$ for $i=0,1,2,...,k-1$. The thin horizontal dotted line indicates the 95\% confidence level to compare with $v_0(x)$.}
  \label{fig:qv}
\end{figure}

Figure \ref{fig:qv} illustrates the $v_0(x)$ and $q_i^*(x)$ returned from algorithm \ref{alg:v} for various $k$. In general, $q_i^*(x)$ is decreasing over $x$, indicating that the optimal proportion invested in the S\&P Composite Index decreases as $c_0$ increases. 

\begin{figure}
  \includegraphics[width=\linewidth]{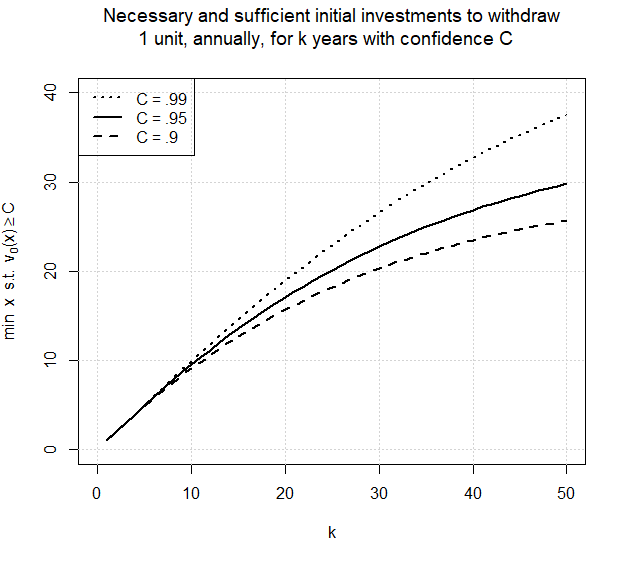}
  \caption{Using algorithm \ref{alg:v} with $w=r=0$, illustrates the minimum $x$ such that $v_0(x)\geq C$ for various $C$ and $k$. Note the assumption that $X_{1i}\sim\mathcal{N}(1.083,.1753^2)$ and $c_{i+1}=-1$ for $i=0,1,2,...,k-1$.}
  \label{fig:ci}
\end{figure}

Figure \ref{fig:ci} illustrates the minimum $x$ such that $v_0(x)\geq C$ for various $C$ and $k$. Of note, an initial investment of $30$ (or $20$) units allows an investor to make $50$ (or $25$) annual withdrawals of $1$ unit with 95\% confidence. 

\begin{figure}
  \includegraphics[width=\linewidth]{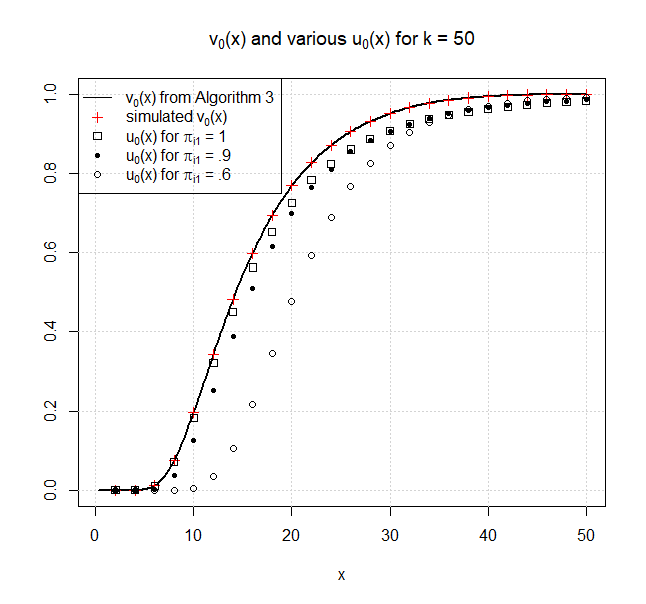}
  \caption{Using algorithm \ref{alg:v} with $M=300$ and $w=r=0$, illustrates $v_0(x)$ for $k=50$. Note the assumption that $X_{1i}\sim\mathcal{N}(1.083,.1753^2)$ and $c_{i+1}=-1$ for $i=0,1,2,...,k-1$. Simulated $v_0(x)$ indicate \eqref{eq:argmax}, computed as described in section \ref{s:lininterp}. The $u_0(x)$ indicate the returned $\mathbb{P}(W_k\geq0)$ from algorithm \ref{alg:usim}, with $c_0=x$ and $\boldsymbol\pi_i=(\pi_{i1},1-\pi_{i1})$ constant over $i$.}
  \label{fig:vcheck}
\end{figure}

Figure \ref{fig:vcheck} shows that the $v_0(x)$ returned from algorithm \ref{alg:v} are reproducable via the Monte Carlo simulation described in section \ref{s:lininterp}. Furthermore, the optimal portfolio weights coming from algorithm \ref{alg:v} offer a noticeable improvement to $\mathbb{P}(W_k\geq 0)$ over constant portfolio weights.

\begin{figure}
  \includegraphics[width=\linewidth]{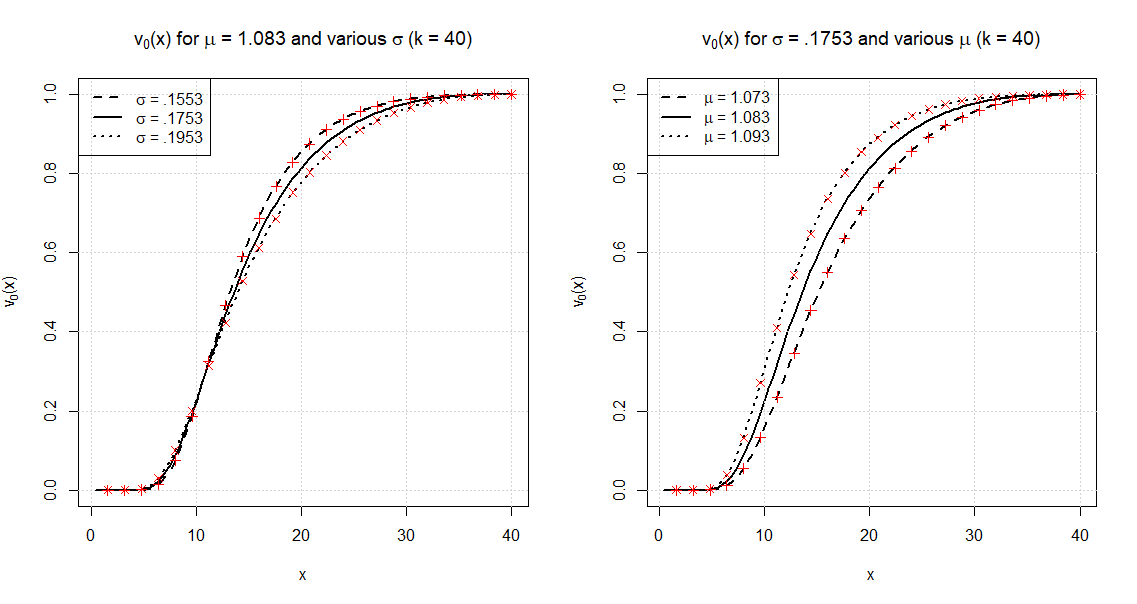}
  \caption{Using algorithm \ref{alg:v} with $M=300$ and $w=r=0$, illustrates $v_0(x)$ for $k=40$ and various combinations of $\mu$ and $\sigma$. Note the assumption that $X_{1i}\sim\mathcal{N}(\mu,\sigma^2)$ and $c_{i+1}=-1$ for $i=0,1,2,...,k-1$. Let $\boldsymbol\pi_i^*$ denote the $\boldsymbol\pi_i$ returned from algorithm \ref{alg:v} using $X_{1i}\sim\mathcal{N}(1.083,.1753^2)$. Left: The red + and $\times$ indicate simulated versions of $\mathbb{P}(W_k\geq0)$ (from algorithm \ref{alg:usim}) when $\boldsymbol\pi_i=\boldsymbol\pi_i^*$ and $X_{1i}\sim\mathcal{N}(\mu,\sigma^2)$ with $(\mu,\sigma)=(1.083,.1553)$ and $(\mu,\sigma)=(1.083,.1953)$, respectively. Right: The red $\times$ and + indicate simulated versions of $\mathbb{P}(W_k\geq0)$ (from algorithm \ref{alg:usim}) when $\boldsymbol\pi_i=\boldsymbol\pi_i^*$ and $X_{1i}\sim\mathcal{N}(\mu,\sigma^2)$ with $(\mu,\sigma)=(1.093,.1553)$ and $(\mu,\sigma)=(1.073,.1753)$, respectively.}
  \label{fig:ms}
\end{figure}

Figure \ref{fig:ms} shows how $v_0(x)$ is affected by $\mu$ and $\sigma$ when $X_{1i}\sim\mathcal{N}(\mu,\sigma^2)$. In general, small changes to $\mu$ and $\sigma$ result in noticeable changes to $v_0(x)$. Let $\boldsymbol\pi_i^*$ denote the $\boldsymbol\pi_i$ returned from algorithm \ref{alg:v} using $X_{1i}\sim\mathcal{N}(1.083,.1753^2)$. Interestingly, the $\mathbb{P}(W_k\geq0)$ returned from algorithm \ref{alg:usim} with $\boldsymbol\pi_i=\boldsymbol\pi_i^*$ and $X_{1i}\sim\mathcal{N}(\mu,\sigma^2)$ are very close to the $v_0(c_0)$ that use $X_{1i}\sim\mathcal{N}(\mu,\sigma^2)$. In other words, the optimal portfolio weights returned from algorithm \ref{alg:v} with $X_{1i}\sim\mathcal{N}(1.083,.1753^2)$ can serve as the optimal portfolio weights when instead $X_{1i}\sim\mathcal{N}(\mu,\sigma^2)$, provided $\mu$ and $\sigma$ are close to $1.083$ and $.1753$, respectively.

Next consider the case where a mortality distribution is not present, and a constant annual withdrawal of 1 unit is made for $k_2$ years, after executing DCA for $k_1$ years ($c_0=c_1=...=c_{k_1-1}>0$ and $c_i=-1$ for $i=k_1,k_1+1,...,k_1+k_2-1$). Table \ref{t:95} shows the DCA investment amount ($c_0$) that supports execution of the $k_2$ years of withdrawals with 95\% confidence. For comparison, the numbers in parenthesis indicate $\mathbb{P}(W_k\geq0)$ ($k=k_1+k_2-1$) when $\boldsymbol\pi_i=(1,0)$ for $i=0,1,...,k-1$, in which case the portfolio is always fully invested in the S\&P Composite Index. In general, the advantage of using optimal portfolio weights instead of $\boldsymbol\pi_i=(1,0)$ decreases as $k_1$ and $k_2$ increase. For the $k_1$ and $k_2$ considered in Table \ref{t:95}, the improvement in $\mathbb{P}(W_k\geq0)$ resulting from using optimal portfolio weights ranges from $.02$ to $.05$. 

\begin{table}
\begin{center}
\caption{Necessary and sufficient $x$ to complete the following schedule with 95\% confidence. Invest $x$ annually for 10, 20, ..., 50 years. Then withdrawal 1 unit annually for 30, 40, ..., 70 years. The $x$ were found by trial and error with algorithm \ref{alg:v} and then checked using algorithm \ref{alg:usim} with $N=100,000$ (as described in section \ref{s:lininterp}). For comparison, the numbers in parenthesis indicate $\mathbb{P}(W_k\geq0)$ (from algorithm \ref{alg:usim}) when $\boldsymbol\pi_i=(1,0)$ for all $i$. Note that $k=k_1+k_2-1$ where $k_1$ is the years of DCA and $k_2$ is the years of withdrawals. Algorithms use $w=r=0$, $M=300$, $N=100,000$ and $X_{1i}\sim\mathcal{N}(1.083,.1753^2)$ for $i=0,1,...,k-1$.}\label{t:95}
\begin{tabular}{ crrrrrr } 
\toprule
 & \multicolumn{5}{c}{Years of Withdrawals} \\
\toprule
Years of DCA &30&40&50&60&70\\
\toprule
10&1.89 (.896)&2.21 (.906)&2.44 (.913)&2.60 (.919)&2.70 (.922)\\ 
20&.76 (.906)&.89 (.916)&.97 (.921)&1.03 (.924)& \\ 
30&.39 (.911)&.46 (.921)&.50 (.924)& & \\ 
40&.23 (.922)&.26 (.924)& & & \\
50&.14 (.930)& & & & \\
\toprule
\end{tabular}
\end{center}
\end{table}

Next consider the case where a mortality distribution is present, and a constant annual withdrawal of 1 unit is made until death, after executing DCA for $k_1$ years ($c_0=c_1=...=c_{k_1-1}>0$ and $c_i=-1$ for $i=k_1,k_1+1,...$). Table \ref{t:95tau} shows the DCA investment amount ($c_0$) that supports execution of the withdrawals until death with 95\% confidence. For comparison, the numbers in parenthesis indicate $\mathbb{P}(\overline{W}_k\geq0)$ when $\boldsymbol\pi_i=(1,0)$ for all $i$, in which case the portfolio is always fully invested in the S\&P Composite Index. In general, the advantage of using optimal portfolio weights instead of $\boldsymbol\pi_i=(1,0)$ decreases as the starting age and $k_1$ increase. For the starting ages and $k_1$ considered in Table \ref{t:95tau}, the improvement in $\mathbb{P}(\overline{W}_k\geq0)$ resulting from using optimal portfolio weights ranges from $.005$ to $.02$. 

\begin{table}
\begin{center}
\caption{Necessary and sufficient $x$ to complete the following schedule with 95\% confidence. Invest $x$ annually for 10, 20, ..., 50 years. Then withdrawal 1 unit annually until death. The $x$ were found by trial and error with algorithm \ref{alg:vhat} and then checked using algorithm \ref{alg:uhatsim} with $N=100,000$ (as described in section \ref{s:lininterp}). For comparison, the numbers in parenthesis indicate $\mathbb{P}(\overline{W}_k\geq0)$ (from algorithm \ref{alg:uhatsim}) when $\boldsymbol\pi_i=(1,0)$ for all $i$. Note that algorithms use $w=r=0$, $M=300$, $N=100,000$ and $X_{1i}\sim\mathcal{N}(1.083,.1753^2)$ for $i=0,1,...,k-1$.}\label{t:95tau}
\begin{tabular}{ crrrrr } 
\toprule
 & \multicolumn{5}{c}{Starting age} \\
\toprule
Years of DCA &20&30&40&50&60\\
\toprule
10&2.58 (.929)&2.42 (.928)&2.19 (.929)&1.91 (.932)&1.54 (.938)\\ 
20&.95 (.930)&.86 (.931)&.75 (.934)&.60 (.940)&\\ 
30&.45 (.936)&.38 (.936)&.30 (.939)&&\\ 
40&.22 (.941)&.17 (.942)&&&\\
50&.10 (.945)&&&&\\
\toprule
\end{tabular}
\end{center}
\end{table}

Last consider the case where a mortality distribution is present, and a constant annual withdrawal of 1 unit is made until death, after an initial investment ($c_0>0$ and $c_i=-1$ for $i=1,2,...$). 

\begin{figure}
  \includegraphics[width=\linewidth]{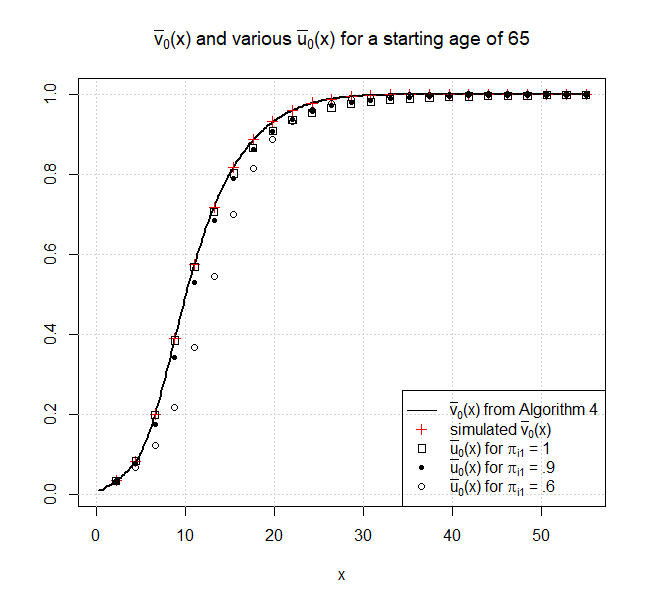}
  \caption{Using algorithm \ref{alg:vhat} with $M=300$ and $w=r=0$, illustrates $\overline{v}_0(x)$ for a starting age of $65$. Note the assumption that $X_{1i}\sim\mathcal{N}(1.083,.1753^2)$ and $c_{i+1}=-1$ for $i=0,1,...$. Simulated $\overline{v}_0(x)$ indicate the right side of \eqref{eq:maxWbar}, computed as described in section \ref{s:lininterp}. The $\overline{u}_0(x)$ indicate the returned $\mathbb{P}(\overline{W}_k\geq0)$ from algorithm \ref{alg:uhatsim}, with $c_0=x$ and $\boldsymbol\pi_i=(\pi_{i1},1-\pi_{i1})$ constant over $i$.}
  \label{fig:s65}
\end{figure}

Figure \ref{fig:s65} shows that the $\overline{v}_0(x)$ returned from algorithm \ref{alg:vhat} are reproducable via the Monte Carlo simulation described in section \ref{s:lininterp}. For $c_0$ between $15$ and $25$, the optimal portfolio weights coming from algorithm \ref{alg:vhat} offer a small, but noticeable, improvement to $\mathbb{P}(\overline{W}_k\geq 0)$ over constant portfolio weights. For other $c_0$, the improvement is hardly distinguishable.

\begin{figure}
  \includegraphics[width=\linewidth]{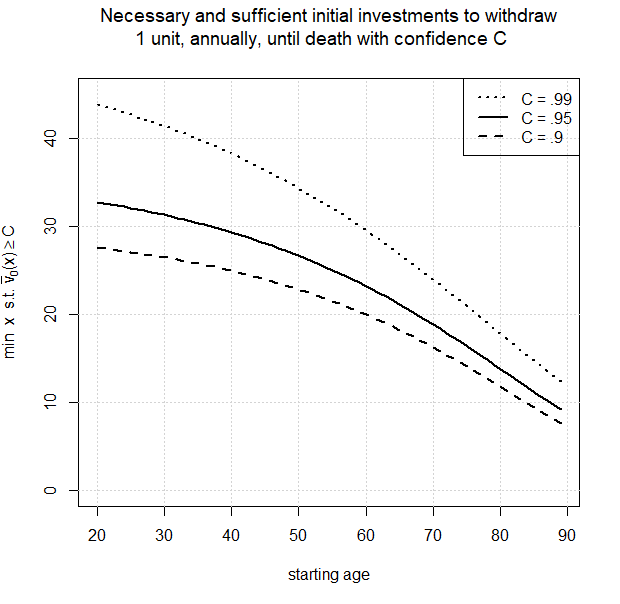}
  \caption{Using algorithm \ref{alg:vhat} with $w=r=0$, illustrates the minimum $x$ such that $\overline{v}_0(x)\geq C$ for various $C$ and $k$. Note the assumption that $X_{1i}\sim\mathcal{N}(1.083,.1753^2)$ and $c_{i+1}=-1$ for $i=0,1,...$.}
  \label{fig:citau}
\end{figure}

Figure \ref{fig:citau} illustrates the minimum $x$ such that $\overline{v}_0(x)\geq C$ for various $C$. Of note, an initial investment of $30$ (or $20$) units at age 60 allows an investor to make annual withdrawals of $1$ unit until death with 99\% (or 90\%) confidence. 

\section{Conclusion}\label{s:conclusion}
In general, maximization of $\mathbb{P}(W_k\geq0)$ and $\mathbb{P}(\overline{W}_k\geq0)$ over the $\boldsymbol\pi_i$ offers a noticeable, but sometimes small, improvement over the $\mathbb{P}(W_k\geq0)$ and $\mathbb{P}(\overline{W}_k\geq0)$ produced when the $\boldsymbol\pi_i$ are constant over $i$. Of the constant portfolio weights, $\boldsymbol\pi_i=(1,0)$ appears to produce $\mathbb{P}(W_k\geq0)$ and $\mathbb{P}(\overline{W}_k\geq0)$ that are closest to their maximum versions; however, the difference can be several tenths. So optimal portfolio weights can offer investors a worthwhile improvement in their probability to complete a given schedule of investments and withdrawals.

This kind of optimization can also be used in the context of a guaranteed lifetime withdrawal benefit (GLWB) rider on a variable annuity. In particular, the GLWB rider that precludes any withdrawals outside of the pre-determined $c_i=-1$ ($i=1,2,...$) can be priced as follows. Consider the recursion
\begin{equation*}
\begin{split}
\widetilde{W}_0&=c_0,\\
\widetilde{W}_i&=\big[\mathcal{X}_{(0,\infty)}(\widetilde{W}_{i-1})Y_{i-1}+1-\mathcal{X}_{(0,\infty)}(\widetilde{W}_{i-1})\big]\widetilde{W}_{i-1}\\
&\quad+(1-\mathcal{X}_{(0,t_i]}(\tau))c_i,\quad i=1,2,...,k.
\end{split}
\end{equation*}
Compute $f:(0,\infty)\to\mathbb{R}$ as
\begin{equation*}
f(c_0)=\underset{\boldsymbol\pi_0,\boldsymbol\pi_1,...,\boldsymbol\pi_{k-1}}{\sup}\mathbb{E}[\widetilde{W}_k].
\end{equation*}
Provided enough customers, a price of $c_0$ will let the insurance company generate a profit of approximately $N\cdot f(c_0)$ at time $t_k$, where $N$ is the number of customers. Moreover, if $t_i$ indicates year $i$ for $i=0,1,...,k$, then the annual withdrawal percentage of the GLWB is given by $(100/c_0)\%$. Typically, variable annuities do not offer this sort of optimized investment, but it could offer customers a noticeable price reduction without changing the annual withdrawal amount. Performing this optimization in the inflation-adjusted setting is especially appealing, since then withdrawals and expected profits will be inflation adjusted. Fortunately, this scheme is easily modified to accommodate a guaranteed death benefit (GDB) rider. If the guaranteed death benefit is $z$ then the expected profit per customer (at time $t_k$) for the variable annuity with GLWB and GDB riders is $f(c_0)-z$. The author plans to investigate this optimization and compare the resulting prices with current variable annuity prices. 

Future research could consider investment and withdrawal amounts $c_i$ that are not pre-determined constants at time 0. At the very least, each $c_i:\Omega\to\mathbb{R}$ would need to be $\mathcal{F}(t_i)$-measurable. On its own, this basic measurability condition leads to trivial optimization. The optimization is more interesting under the following scheme. Fix $k\in\mathbb{N}$ and $C\in(0,1)$. At each time $t_i$, compute $g:\mathbb{R}\to[0,1]$ such that $g(x)=v_i(W_i)$, where $v_i(W_i)$ is computed using $c_j=x$ for $j=i,i+1,...,k$. Set $c_i$ to be the least $x$ such that $g(x)\geq C$. If there is no such $x$, set $c_i=0$. In words this scheme starts with an initial investment of $c_0>0$, and then the $c_i$ are non-positive for $i=1,2,...,k$, with each $c_i$ being adapted in time to maximize the probability (given the information up to time $t_i$) of withdrawing $-c_i$ at each time $t_i,t_{i+1},...,t_k$. It would be interesting to investigate the distribution of these adapted $c_i$, meaning look at $\mathbb{P}(c_i\leq -y)$ for a desirable withdrawal amount $y>0$. Ideally, the $\mathbb{P}(c_i\leq -y)$ should be greater than or equal to some threshold $C\in(0,1)$. Ultimately, the goal of this scheme would be to achieve increased $\mathbb{P}(W_k\geq0)$ at the cost of having variable withdrawal amounts. 

\begin{appendices}

\section{Proofs}\label{secA1}

\subsection{Additional notation and definitions}
The proofs in section \ref{secA1} involve the following notation and definitions, which were not mentionded in section \ref{s:notation}. Use $\mathbb{R}^*$ to denote the extended real numbers. If $X$ and $Y$ are sets and $f:X\to Y$, the graph of $f$ is
\begin{equation*}
\text{Gr}(f)=\{(x,f(x))\|\ x\in X\}.
\end{equation*}
Furthermore, the $x$-section of $D\subset X\times Y$ is given by
\begin{equation*}
D_x=\{y:(x,y)\in D\},
\end{equation*}
and the projection of $D$ onto $X$ is given by
\begin{equation*}
proj_X(D)=\{x:(x,y)\in D\}.
\end{equation*}

\subsection{Additional lemmas}
The proofs in section \ref{secA1} involve the following lemmas, which were not mentionded in section \ref{s:notation}.

\begin{lemma}[simplified version of proposition 7.30 from \cite{bertsekas1996stochastic}]\label{l:cont}
Let $X$ and $Y$ be separable metrizable spaces and let $\mu$ be a Borel probability measure on the Borel $\sigma$-algebra of $Y$. If $f:X\times Y\to\mathbb{R}$ is bounded and continuous, then the function $\lambda:X\to\mathbb{R}$ defined by
\begin{equation*}
\lambda(x)=\int_Yf(x,y)d\mu(y)
\end{equation*}
is continuous.
\end{lemma}
\begin{lemma}[simplified version of proposition 7.31 from \cite{bertsekas1996stochastic}]\label{l:semi}
Let $X$ and $Y$ be separable metrizable spaces, let $\mu$ be a Borel probability measure on the Borel $\sigma$-algebra of $Y$, and let $f:X\times Y\to\mathbb{R}$ be Borel measurable. If $f$ is upper semicontinuous and bounded above, the so is the function $\lambda:X\to\mathbb{R}$ defined by
\begin{equation*}
\lambda(x)=\int_Yf(x,y)d\mu(y).
\end{equation*}
\end{lemma}
\begin{lemma}[proposition 7.33 of \cite{bertsekas1996stochastic}] \label{l:sup}
Let $X$ be a metrizable space, $Y$ a compact metrizable space, $D$ a closed subset of $X\times Y$, and let $f:D\to\mathbb{R}^*$ be upper semicontinuous. Let $f^*:\text{proj}_X(D)\to\mathbb{R}^*$ be given by
\begin{equation*}
f^*(x)=\sup_{y\in D_x}f(x,y).
\end{equation*}
Then $\text{proj}_X(D)$ is closed in $X$, $f^*$ is upper semicontinuous, and there exists a Borel measurable function $\psi:\text{proj}_X(D)\to Y$ such that $\text{Gr}(\psi)\subset D$ and 
\begin{equation*}
f(x,\psi(x))=f^*(x)\quad\forall  x\in\text{proj}_X(D).
\end{equation*}
\end{lemma}

\subsection{Proof that $v_i$ are well-defined}\label{s:Av}
Obviously, $v_k$ is well-defined. It remains to be seen whether the $v_i$ are well-defined for $i=0,1,...,k-1$. To begin, $v_i$ is given in another representation that \eqref{eq:VI} will be shown to satisfy. Let $v_i:\mathbb{R}\to[0,1]$, $i=0,1,...,k-1$, denote the non-decreasing upper semicontinuous functions satisfying
\begin{equation}\label{eq:vi}
v_i(W_i)=\sup_{\boldsymbol\pi_i}\mathbb{E}[v_{i+1}(W_{i+1})\ \vert\ W_i]\quad\text{a.s.}
\end{equation}
Note that the supremum in \eqref{eq:vi} indicates the supremum over all $\mathcal{F}(t_i)$-measurable portfolio weight functions $\boldsymbol\pi_i$. 

In order for the recursion given by \eqref{eq:vi} to make sense, the $v_i$ need to be well-defined. This is shown via induction. Suppose that $v_{i+1}$ is well-defined. The goal from here is to show that $v_i$ is well-defined.

Observe that $v_{i+1}(W_{i+1})$ is $\mathcal{F}$-measurable and integrable because $v_{i+1}$ is bounded and Borel measurable, and $W_{i+1}$ is $\mathcal{F}$-measurable. So $\mathbb{E}[v_{i+1}(W_{i+1})\ \vert\ W_i]$ is well-defined.

Since $\sigma(W_i)\subset\mathcal{F}(t_i)$, a property of conditional expectation gives 
\begin{equation}\label{eq:evevev}
\mathbb{E}[v_{i+1}(W_{i+1})\ \vert\ W_i]=\mathbb{E}[\mathbb{E}[v_{i+1}(W_{i+1})\ \vert\ \mathcal{F}(t_i)]\ \vert\ W_i] \quad\text{a.s.}
\end{equation}
Recall from Section \ref{s:notation} and \eqref{recursion} that $W_{i+1}=(\boldsymbol\pi_{i}\cdot\mathbf{X}_{i})W_i+c_{i+1}$. Observe that $\boldsymbol\pi_i$ and $W_i$ are both $\mathcal{F}(t_i)$-measurable, and $\mathbf{X}_i$ is independent of $\mathcal{F}(t_i)$. By lemma \ref{l:ce},
\begin{equation}\label{eq:evH}
\mathbb{E}[v_{i+1}(W_{i+1})\ \vert\ \mathcal{F}(t_i)]=H(\boldsymbol\pi_i,W_i) \quad\text{a.s.},
\end{equation}
where 
\begin{equation}\label{eq:Hdef}
H(\mathbf{p},x)=\mathbb{E}[v_{i+1}((\mathbf{p}\cdot\mathbf{X}_{i})x+c_{i+1})]
\end{equation}
for each $\mathbf{p}\in\Pi$ and $x\in\mathbb{R}$. It follows from \eqref{eq:evevev} and \eqref{eq:evH} that
\begin{equation}\label{eq:eveH}
\mathbb{E}[v_{i+1}(W_{i+1})\ \vert\ W_i]=\mathbb{E}[H(\boldsymbol\pi_i,W_i)\ \vert\ W_i] \quad\text{a.s.}
\end{equation}
The next goal is to show that 
\begin{equation}\label{eq:supeH}
\sup_{\boldsymbol\pi_i}\mathbb{E}[H(\boldsymbol\pi_i,W_i)\ \vert\ W_i]=\mathbb{E}[\sup_{\boldsymbol\pi_i}H(\boldsymbol\pi_i,W_i)\ \vert\ W_i]. \quad\text{a.s.}
\end{equation}
Observe that 
\begin{equation*}
H(\boldsymbol\pi,W_i)\leq\sup_{\boldsymbol\pi_i}H(\boldsymbol\pi_i,W_i)\quad\text{a.s.}
\end{equation*}
for each $\mathcal{F}(t_i)$-measurable portfolio weight function $\boldsymbol\pi$. It follows that
\begin{equation*}
\mathbb{E}[H(\boldsymbol\pi,W_i)\ \vert\ W_i]\leq\mathbb{E}[\sup_{\boldsymbol\pi_i}H(\boldsymbol\pi_i,W_i)\ \vert\ W_i]\quad\text{a.s.}
\end{equation*}
So \eqref{eq:supeH} holds when $=$ is replaced with $\leq$. To show that \eqref{eq:supeH} holds when $=$ is replaced with $\geq$, first observe that 
 \begin{equation}\label{eq:supeHgeq}
\begin{split}
\sup_{\boldsymbol\pi_i}\mathbb{E}[H(\boldsymbol\pi_i,W_i)\ \vert\ W_i]&\geq\sup_{\boldsymbol\phi}\mathbb{E}[H(\boldsymbol\phi(W_i),W_i)\ \vert\ W_i]\\
&=\sup_{\boldsymbol\phi}H(\boldsymbol\phi(W_i),W_i) \quad\text{a.s.},
\end{split}
\end{equation}
where the supremum is taken over the Borel measurable functions $\boldsymbol\phi:\mathbb{R}\to\Pi$. Because of \eqref{eq:supeHgeq}, it suffices to show
\begin{equation}\label{eq:shehw}
\sup_{\boldsymbol\phi}H(\boldsymbol\phi(W_i),W_i)=\mathbb{E}[\sup_{\boldsymbol\pi_i}H(\boldsymbol\pi_i,W_i)\ \vert\ W_i]. \quad\text{a.s.}
\end{equation}
This is done by showing 
\begin{equation}\label{eq:supH}
\sup_{\boldsymbol\pi_i}H(\boldsymbol\pi_i,W_i)=\sup_{\boldsymbol\phi}H(\boldsymbol\phi(W_i),W_i)\quad\text{a.s.},
\end{equation}
and
\begin{equation}\label{eq:supH2}
\mathbb{E}[\sup_{\boldsymbol\phi}H(\boldsymbol\phi(W_i),W_i)\ \vert\ W_i]=\sup_{\boldsymbol\phi}H(\boldsymbol\phi(W_i),W_i)\quad\text{a.s.}
\end{equation}

The $\boldsymbol\phi(W_i)$ in \eqref{eq:supH} are $\sigma(W_i)$-measurable. Since $\sigma(W_i)\subset\mathcal{F}(t_i)$, it follows that \eqref{eq:supH} holds when $=$ is replaced with $\geq$. To show that \eqref{eq:supH} holds when $=$ is replaced with $\leq$, observe that for almost every $\omega\in\Omega$,
\begin{equation}\label{eq:shphishr}
\begin{split}
(\sup_{\boldsymbol\pi_i}H(\boldsymbol\pi_i,W_i))(\omega)&=\sup_{\boldsymbol\pi_i}H(\boldsymbol\pi_i(\omega),W_i(\omega))\\
&\leq\sup_{\mathbf{r}\in\Pi}H(\mathbf{r},W_i(\omega))\\
&\leq\sup_{\boldsymbol\phi}H(\boldsymbol\phi(W_i(\omega)),W_i(\omega))\\
&=(\sup_{\boldsymbol\phi}H(\boldsymbol\phi(W_i),W_i))(\omega).
\end{split}
\end{equation}
Moreover, it is now clear from \eqref{eq:shphishr} that
\begin{equation}\label{eq:phitor}
\sup_{\boldsymbol\phi}H(\boldsymbol\phi(W_i),W_i)=\sup_{\mathbf{r}\in\Pi}H(\mathbf{r},W_i)\quad\text{a.s.}
\end{equation}
To show \eqref{eq:supH2}, it suffices to show that the right side of \eqref{eq:supH2} is $\sigma(W_i)$-measurable. Let $G:\Pi\times\mathbb{R}\times(0,\infty)^n$ be such that
\begin{equation*}
G(\mathbf{r},x,\mathbf{X})=(\mathbf{r}\cdot\mathbf{X})x+c_{i+1}.
\end{equation*}
Observe that $G$ is continuous. Since $v_{i+1}$ is upper semicontinuous, it follows that $v_{i+1}\circ G$ is upper semicontinuous. Applying lemma \ref{l:semi} reveals that $H$ is upper semicontinuous. To see this, let $f$, $X$ and $Y$ in lemma \ref{l:semi} be $v_{i+1}\circ G$, $\Pi\times\mathbb{R}$ and $(0,\infty)^n$, respectively. Next, applying lemma \ref{l:sup} reveals that $\sup_{\mathbf{r}\in\Pi}H(\mathbf{r},x)$ is upper semicontinuous over $x\in\mathbb{R}$, and there is a Borel measurable function $\boldsymbol\phi^*:\mathbb{R}\to\Pi$ such that 
\begin{equation}\label{eq:phieqsup}
H(\boldsymbol\phi^*(x),x)=\sup_{\mathbf{r}\in\Pi}H(\mathbf{r},x)\quad\forall x\in\mathbb{R}.
\end{equation}
Therefore $\sup_{\mathbf{r}\in\Pi}H(\mathbf{r},x)$ is Borel measurable over $x\in\mathbb{R}$, which implies $\sup_{\mathbf{r}\in\Pi}H(\mathbf{r},W_i)$ is $\sigma(W_i)$-measurable. Applying \eqref{eq:phitor}, the right side of \eqref{eq:supH2} is $\sigma(W_i)$-measurable. Now it follows from \eqref{eq:vi}, \eqref{eq:eveH}, \eqref{eq:supeH}, \eqref{eq:shehw}, \eqref{eq:phitor} and \eqref{eq:phieqsup} that 
\begin{equation}\label{eq:viH}
v_i(W_i)=\sup_{\mathbf{r}\in\Pi}H(\mathbf{r},W_i)=H(\boldsymbol\phi^*(W_i),W_i)\quad\text{a.s.}
\end{equation}
Looking at \eqref{eq:Hdef}, it is not hard to see that $\sup_{\mathbf{r}\in\Pi}H(\mathbf{r},x)$ is non-decreasing over $x\in\mathbb{R}$. Furthermore, the fact that $v_{i+1}$ maps to $[0,1]$ implies $\sup_{\mathbf{r}\in\Pi}H(\mathbf{r},x)$ also maps to $[0,1]$. It is now clear that the $v_i$ as given in \eqref{eq:VI} are well-defined and satisfy \eqref{eq:vi}. 

\subsection{Proof that \eqref{eq:argmax} is given by $v_0(c_0)$}\label{s:Av0}
Observe that
\begin{equation}\label{eq:etoev}
\mathbb{P}(W_k\geq w)=\mathbb{E}[v_k(W_k)].
\end{equation}
By the law of total expectation, for each $i=0,1,...,k-1$, 
\begin{equation}\label{eq:supte}
\mathbb{E}[v_{i+1}(W_{i+1})]=\mathbb{E}[\mathbb{E}[v_{i+1}(W_{i+1})\ \vert\ W_i]].
\end{equation}

Next, the goal is to show
\begin{equation}\label{eq:supoutsupin}
\underset{\boldsymbol\pi_0,\boldsymbol\pi_1,...,\boldsymbol\pi_{i}}{\sup}\mathbb{E}[\mathbb{E}[v_{i+1}(W_{i+1})\ \vert\ W_i]]=\underset{\boldsymbol\pi_0,\boldsymbol\pi_1,...,\boldsymbol\pi_{i-1}}{\sup}\mathbb{E}[\sup_{\boldsymbol\pi_{i}}\mathbb{E}[v_{i+1}(W_{i+1})\ \vert\ W_i]].
\end{equation}
Obviously,
\begin{equation}\label{eq:evwsevw}
\mathbb{E}[v_{i+1}(W_{i+1})\ \vert\ W_i]\leq\sup_{\boldsymbol\pi_{i}}\mathbb{E}[v_{i+1}(W_{i+1})\ \vert\ W_i]\quad\text{a.s.}
\end{equation}
It follows from \eqref{eq:evwsevw} that \eqref{eq:supoutsupin} holds when $=$ is replaced with $\leq$. To show that \eqref{eq:supoutsupin} holds when $=$ is replaced with $\geq$, first recall from \eqref{eq:eveH} that
\begin{equation*}
\mathbb{E}[v_{i+1}(W_{i+1})\ \vert\ W_i]=\mathbb{E}[H(\boldsymbol\pi_i,W_i)\ \vert\ W_i]\quad\text{a.s.}
\end{equation*}
Therefore
\begin{equation*}
\begin{split}
\underset{\boldsymbol\pi_0,\boldsymbol\pi_1,...,\boldsymbol\pi_{i}}{\sup}\mathbb{E}[\mathbb{E}[v_{i+1}(W_{i+1})\ \vert\ W_i]]&\geq\underset{\boldsymbol\pi_0,\boldsymbol\pi_1,...,\boldsymbol\pi_{i}=\boldsymbol\phi(W_i)}{\sup}\mathbb{E}[\mathbb{E}[v_{i+1}(W_{i+1})\ \vert\ W_i]]\\
&=\underset{\boldsymbol\pi_0,\boldsymbol\pi_1,...,\boldsymbol\pi_{i}=\boldsymbol\phi(W_i)}{\sup}\mathbb{E}[\mathbb{E}[H(\boldsymbol\phi(W_i),W_i)\ \vert\ W_i]]\\
&=\underset{\boldsymbol\pi_0,\boldsymbol\pi_1,...,\boldsymbol\pi_{i}=\boldsymbol\phi(W_i)}{\sup}\mathbb{E}[H(\boldsymbol\phi(W_i),W_i)]\\
&=\underset{\boldsymbol\pi_0,\boldsymbol\pi_1,...,\boldsymbol\pi_{i-1},\boldsymbol\phi}{\sup}\mathbb{E}[H(\boldsymbol\phi(W_i),W_i)].
\end{split}
\end{equation*}
Next observe that \eqref{eq:vi} and \eqref{eq:viH} imply
\begin{equation*}
\sup_{\boldsymbol\pi_{i}}\mathbb{E}[v_{i+1}(W_{i+1})\ \vert\ W_i]=H(\boldsymbol\phi^*(W_i),W_i)\quad\text{a.s.}
\end{equation*}
So now it suffices to show
\begin{equation}\label{eq:eHgeq}
\underset{\boldsymbol\pi_0,\boldsymbol\pi_1,...,\boldsymbol\pi_{i-1},\boldsymbol\phi}{\sup}\mathbb{E}[H(\boldsymbol\phi(W_i),W_i)]\geq\underset{\boldsymbol\pi_0,\boldsymbol\pi_1,...,\boldsymbol\pi_{i-1}}{\sup}\mathbb{E}[H(\boldsymbol\phi^*(W_i),W_i)],
\end{equation}
which is clearly true.

Combining \eqref{eq:supte}, \eqref{eq:supoutsupin} and \eqref{eq:vi},
\begin{equation}\label{eq:suppik}
\begin{split}
\underset{\boldsymbol\pi_0,\boldsymbol\pi_1,...,\boldsymbol\pi_{i}}{\sup}\mathbb{E}[v_{i+1}(W_{i+1})]&=\underset{\boldsymbol\pi_0,\boldsymbol\pi_1,...,\boldsymbol\pi_{i}}{\sup}\mathbb{E}[\mathbb{E}[v_{i+1}(W_{i+1})\ \vert\ W_i]]\\
&=\underset{\boldsymbol\pi_0,\boldsymbol\pi_1,...,\boldsymbol\pi_{i-1}}{\sup}\mathbb{E}[\sup_{\boldsymbol\pi_{i}}\mathbb{E}[v_{i+1}(W_{i+1})\ \vert\ W_i]]\\
&=\underset{\boldsymbol\pi_0,\boldsymbol\pi_1,...,\boldsymbol\pi_{i-1}}{\sup}\mathbb{E}[v_i(W_i)].
\end{split}
\end{equation}
By repeated application of \eqref{eq:suppik},
\begin{equation}\label{eq:vce}
\begin{split}
\underset{\boldsymbol\pi_0,\boldsymbol\pi_1,...,\boldsymbol\pi_{k-1}}{\sup}\mathbb{E}[v_k(W_k)]&=\underset{\boldsymbol\pi_0,\boldsymbol\pi_1,...,\boldsymbol\pi_{k-2}}{\sup}\mathbb{E}[v_{k-1}(W_{k-1})]\\
&=\underset{\boldsymbol\pi_0,\boldsymbol\pi_1,...,\boldsymbol\pi_{k-3}}{\sup}\mathbb{E}[v_{k-2}(W_{k-2})]\\
&=\ .\ .\ .\ \\
&=\mathbb{E}[v_0(W_0)].
\end{split}
\end{equation}
Again, $W_0=c_0$ is deterministic, so $\mathbb{E}[v_0(W_0)]=v_0(c_0)$. It follows from \eqref{eq:etoev} and \eqref{eq:vce} that
\begin{equation}\label{eq:ptov}
\underset{\boldsymbol\pi_0,\boldsymbol\pi_1,...,\boldsymbol\pi_{k-1}}{\sup}\mathbb{P}(W_k\geq w)=v_0(c_0).
\end{equation}

\subsection{Proofs for stock-bond portfolios}\label{s:Asb}
Observe that $v_k$ is the indicator function of $[w_k,\infty)$, so $v_k$ is continuous over $\mathbb{R}\setminus\{w_k\}$ and right continuous at $w_k$. Now proceed by induction. Suppose $v_{i+1}$ is continuous over $\mathbb{R}\setminus\{w_{i+1}\}$ and right continuous at $w_{i+1}$. Further suppose that $v_{i+1}(x)=1$ for $x\geq w_{i+1}$, and if $c_{i+2},...,c_k\leq 0$, suppose that $v_{i+1}(x)=0$ for $x<0$. 

For $\mathbf{r}\in\Pi$ and $x\in\mathbb{R}$, let $A_{\mathbf{r},x}=\{\omega:(\mathbf{r}\cdot\mathbf{X}_i)x+c_{i+1}=w_{i+1}\}$. Using the notation of section \ref{s:Av}, observe that 
\begin{equation*}
H(\mathbf{r},x)=\int_{\Omega\setminus A_{\mathbf{r},x}}(v_{i+1}\circ G)(\mathbf{r},x,\mathbf{X}_i)d\mathbb{P}+\int_{A_{\mathbf{r},x}}(v_{i+1}\circ G)(\mathbf{r},x,\mathbf{X}_i)d\mathbb{P}.
\end{equation*}
Under the setting outlined at the beginning of section \ref{s:sb}, $\mathbb{P}(A_{\mathbf{r},x})=0$ for $(\mathbf{r},x)\neq((0,1),w_i)$. Therefore
\begin{equation*}
H(\mathbf{r},x)=\int_{\Omega\setminus A_{\mathbf{r},x}}(v_{i+1}\circ G)(\mathbf{r},x,\mathbf{X}_i)d\mathbb{P}
\end{equation*}
when $(\mathbf{r},x)\neq((0,1),w_i)$. By assumption, $v_{i+1}$ is continuous over $\mathbb{R}\setminus\{w_{i+1}\}$. Also recall from section \ref{s:Av} that $G$ is continuous. Therefore $v_{i+1}\circ G$ is continuous over 
\begin{equation*}
\{(\mathbf{r},x,\mathbf{X})\in\Pi\times\mathbb{R}\times (0,\infty)^2:(\mathbf{r}\cdot\mathbf{X})x+c_{i+1}\neq w_{i+1}\}.
\end{equation*}
It follows from lemma \ref{l:cont} that $H(\mathbf{r},x)$ is continuous over $(\Pi\times\mathbb{R})\setminus\{((0,1),w_i)\}$. To see that $v_i(x)$ is continuous at any point $x\in(-\infty,w_i)$, simply take advantage of the uniform continuity of $H$ over $\Pi\times[x-1,\ (x+w_i)/2]$. 

Observe that
\begin{equation*}
H((0,1),x)=\mathbb{E}[v_{i+1}((1+r)x+c_{i+1})]=v_{i+1}((1+r)x+c_{i+1}).
\end{equation*}
Since $v_{i+1}(x)=1$ for $x\geq w_{i+1}$, it follows that $H((0,1),x)=1$ for $x\geq w_i$. Recall that $v_i$ maps to $[0,1]$ and $v_i(x)\geq H((0,1),x)$ for each $x\in\mathbb{R}$. Therefore $v_i(x)=1$ for $x\geq w_i$. 

The next goal is to show $v_i(x)=0$ for $x<0$, provided $c_{i+1},...,c_k\leq0$. Since $c_{i+2},...,c_k\leq0$, it follows from the assumption in the first paragraph of this section that $v_{i+1}(x)=0$ for $x<0$. Additionally having $c_{i+1}\leq0$ implies that for each $\mathbf{r}\in\Pi$ and $x<0$, $H(\mathbf{r},x)=0$. Thus, $v_i(x)=0$ for $x<0$.

Showing \eqref{eq:visb} reduces to showing
\begin{equation}\label{eq:vB}
\int_Bv_{i+1}((\mathbf{r}\cdot\mathbf{X}_i)x+c_{i+1})d\mathbb{P}=0
\end{equation}
for $x\geq0$, where $B=\{\omega:(\mathbf{r}\cdot\mathbf{X}_i)x+c_{i+1}<0\}$. If $c_{i+1}\geq0$, then $\mathbb{P}(B)=0$ and \eqref{eq:vB} follows. If $c_{i+1}<0$, then $c_{i+2},...,c_k\leq 0$ as well (see section \ref{s:notation}). By the previous paragraph, $v_{i+1}(x)=0$ for $x<0$. So \eqref{eq:vB} holds when $c_{i+1}<0$.

\subsection{Proof that the right side of \eqref{eq:maxWbar} is given by $\overline{v}_0(c_0)$}\label{s:Avbar}
This proof is very similar to that of sections \ref{s:Av} and \ref{s:Av0}, so only an outline with the key differences is presented. First introduce the additional recursion
\begin{equation}
\begin{split}
\overline{\tau}_0&=0,\\
\overline{\tau}_i&=\overline{\tau}_{i-1}+\tau_{i-1},\quad i=1,2,...,k.
\end{split}
\label{taubar}
\end{equation}
Next observe that for $i=1,2,...,k$,
\begin{equation*}
\mathcal{X}_{(0,t_i]}(\tau)=\mathcal{X}_{[1,\infty)}(\sum_{j=0}^{i-1}\tau_j)=\mathcal{X}_{[1,\infty)}(\overline{\tau}_{i-1}+\tau_{i-1}).
\end{equation*}
Therefore
\begin{equation}
\begin{split}
\overline{W}_i&=(1-\mathcal{X}_{[1,\infty)}(\overline{\tau}_{i-1}+\tau_{i-1}))(Y_{i-1}\overline{W}_{i-1}+c_i)\\
&\quad+\mathcal{X}_{[1,\infty)}(\overline{\tau}_{i-1}+\tau_{i-1})\overline{W}_{i-1},\quad i=1,2,...,k.
\end{split}
\label{Wtaubar}
\end{equation}

Like in section \ref{s:Av}, $\overline{v}_i$ is given first in another representation that \eqref{eq:evbarix} will be shown to satisfy. Let $\widehat{v}_k:\mathbb{R}\times\mathbb{N}_0$ be such that $\widehat{v}_k(x,t)=v_k(x)$, where $\mathbb{N}_0=\{0,1,...\}$. For $i=0,1,...,k-1$, let $\widehat{v}_i:\mathbb{R}\times\mathbb{N}_0\to[0,1]$ denote the functions satisfying
\begin{equation}\label{eq:vbari}
\widehat{v}_i(\overline{W}_i,\overline{\tau}_i)=\sup_{\boldsymbol\pi_i}\mathbb{E}[\widehat{v}_{i+1}(\overline{W}_{i+1},\overline{\tau}_{i+1})\ \vert\ (\overline{W}_i,\overline{\tau}_i)]\quad\text{a.s.},
\end{equation}
with 
\begin{itemize}
\item $\widehat{v}_i(x,t)$ upper semicontinuous w.r.t. $x$ for each $t\in\mathbb{N}_0$,
\item $\widehat{v}_{i}(x,t)=v_k(x)$ for $x\in\mathbb{R}$ and $t=1,2,...$,
\item $\widehat{v}_{i}(x,0)$ non-decreasing and $\widehat{v}_{i}(x,0)=\overline{v}_i(x)$ for $x\in\mathbb{R}$.
\end{itemize}
Suppose that $\widehat{v}_{i+1}$ is well-defined. The goal from here is to show that $\overline{v}_i$ is well-defined. Observe that 
\begin{equation}
\begin{split}
\widehat{v}_{i+1}(\overline{W}_{i+1},\overline{\tau}_{i+1})&=\widehat{v}_{i+1}\Big((1-\mathcal{X}_{[1,\infty)}(\overline{\tau}_i+\tau_i))(Y_{i}\overline{W}_{i}+c_{i+1})\\
&\quad\quad\quad\quad+\mathcal{X}_{[1,\infty)}(\overline{\tau}_i+\tau_i)\overline{W}_{i},\ \overline{\tau}_i+\tau_i\Big)\\
&=(1-\mathcal{X}_{[1,\infty)}(\overline{\tau}_i+\tau_i))\widehat{v}_{i+1}(Y_{i}\overline{W}_{i}+c_{i+1},\ \overline{\tau}_i+\tau_i)\\
&\quad+\mathcal{X}_{[1,\infty)}(\overline{\tau}_i+\tau_i)\widehat{v}_{i+1}(\overline{W}_{i},\ \overline{\tau}_i+\tau_i).
\end{split}
\label{eq:vwtaup1}
\end{equation}
Since $\sigma((\overline{W}_i,\overline{\tau}_i))\subset\mathcal{F}(t_i)$, a property of conditional expectation gives
\begin{equation}\label{eq:vbarcece}
\begin{split}
&\mathbb{E}[\widehat{v}_{i+1}(\overline{W}_{i+1},\overline{\tau}_{i+1})\ \vert\ (\overline{W}_i,\overline{\tau}_i)]\\
&\quad=\mathbb{E}[\mathbb{E}[\widehat{v}_{i+1}(\overline{W}_{i+1},\overline{\tau}_{i+1})\ \vert\ \mathcal{F}(t_i)]\ \vert\ (\overline{W}_i,\overline{\tau}_i)]\quad\text{a.s.}
\end{split}
\end{equation}
It follows from \eqref{eq:vwtaup1} and lemma \ref{l:ce} that 
\begin{equation}\label{eq:evbarH}
\mathbb{E}[\widehat{v}_{i+1}(\overline{W}_{i+1},\overline{\tau}_{i+1})\ \vert\ \mathcal{F}(t_i)]=\overline{H}(\boldsymbol\pi_i,\overline{W}_i,\overline{\tau}_i)\quad\text{a.s.},
\end{equation}
where 
\begin{equation*}
\begin{split}
\overline{H}(\mathbf{p},x,t)&=\mathbb{E}[(1-\mathcal{X}_{[1,\infty)}(t+\tau_i))\widehat{v}_{i+1}((\mathbf{p}\cdot\mathbf{X}_i)x+c_{i+1},\ t+\tau_i)\\
&\quad\quad+\mathcal{X}_{[1,\infty)}(t+\tau_i)\widehat{v}_{i+1}(x,\ t+\tau_i)]
\end{split}
\end{equation*}
for each $\mathbf{p}\in\Pi$, $x\in\mathbb{R}$ and $t\in\mathbb{N}_0$. By the independence of $\tau_i$ to each $\mathbf{X}_i$, it follows that
\begin{equation*}
\overline{H}(\mathbf{p},x,t)=\begin{cases}
(1-p_i)\mathbb{E}[\widehat{v}_{i+1}((\mathbf{p}\cdot\mathbf{X}_i)x+c_{i+1},0)]+p_i\widehat{v}_{i+1}(x,1)&t=0,\\
(1-p_i)\widehat{v}_{i+1}(x,t)+p_i\widehat{v}_{i+1}(x,t+1)&t=1,2,...
\end{cases}
\end{equation*}
By the induction assumption, $\widehat{v}_{i+1}(x,t)=v_k(x)$ and $\widehat{v}_{i+1}(x,0)=\overline{v}_{i+1}(x)$ for $x\in\mathbb{R}$ and $t=1,2,...$. Therefore
\begin{equation*}
\overline{H}(\mathbf{p},x,t)=\begin{cases}
(1-p_i)\mathbb{E}[\overline{v}_{i+1}((\mathbf{p}\cdot\mathbf{X}_i)x+c_{i+1})]+p_iv_k(x)&t=0,\\
v_k(x)&t=1,2,...
\end{cases}
\end{equation*}
It follows from \eqref{eq:vbarcece} and \eqref{eq:evbarH} that
\begin{equation}\label{eq:evbarH2}
\mathbb{E}[\widehat{v}_{i+1}(\overline{W}_{i+1},\overline{\tau}_{i+1})\ \vert\ (\overline{W}_i,\overline{\tau}_i)]=\mathbb{E}[\overline{H}(\boldsymbol\pi_i,\overline{W}_i,\overline{\tau}_i)\ \vert\ (\overline{W}_i,\overline{\tau}_i)]\quad\text{a.s.}
\end{equation}
Paralleling the logic of section \ref{s:Av}, the following result is eventually achieved. $\sup_{\mathbf{r}\in\Pi}\overline{H}(\mathbf{r},x,t)$ is upper semicontinuous over $x$ for each $t\in\mathbb{N}_0$, and there is a Borel measurable function $\overline{\boldsymbol\phi}:\mathbb{R}\times\mathbb{N}_0\to\Pi$ such that 
\begin{equation*}\label{eq:viH}
\widehat{v}_i(\overline{W}_i,\overline{\tau}_i)=\sup_{\mathbf{r}\in\Pi}H(\mathbf{r},\overline{W}_i,\overline{\tau}_i)=\overline{H}(\overline{\boldsymbol\phi}(\overline{W}_i,\overline{\tau}_i),\overline{W}_i,\overline{\tau}_i)\quad\text{a.s.}
\end{equation*}
Moreover, a satisfactory $\widehat{v}_i$ is given by 
\begin{equation*}
\widehat{v}_i(x,t)=\sup_{\mathbf{r}\in\Pi}\overline{H}(\mathbf{r},x,t),\quad x\in\mathbb{R},\ t\in\mathbb{N}_0. 
\end{equation*}
$\widehat{v}_i$ is well-defined because $\overline{H}$ is a well-defined mapping to $[0,1]$ and 
\begin{itemize}
\item $\sup_{\mathbf{r}\in\Pi}\overline{H}(\mathbf{r},x,t)$ is upper semicontinuous w.r.t. $x$ for each $t\in\mathbb{N}_0$,
\item $\sup_{\mathbf{r}\in\Pi}\overline{H}(\mathbf{r},x,t)=v_k(x)$ for $x\in\mathbb{R}$ and $t=1,2,...$ ,
\item $\sup_{\mathbf{r}\in\Pi}\overline{H}(\mathbf{r},x,0)$ is non-decreasing and $\sup_{\mathbf{r}\in\Pi}\overline{H}(\mathbf{r},x,0)=\overline{v}_i(x)$ for $x\in\mathbb{R}$.
\end{itemize}
Likewise, $\overline{v}_i$ is well-defined because $\overline{H}$ is a well-defined mapping to $[0,1]$ and 
\begin{itemize}
\item $\sup_{\mathbf{r}\in\Pi}\overline{H}(\mathbf{r},x,0)$ is upper semicontinuous w.r.t. $x$,
\item $\sup_{\mathbf{r}\in\Pi}\overline{H}(\mathbf{r},x,0)$ is non-decreasing w.r.t. $x$.
\end{itemize}

From here, this proof mimics that of section \ref{s:Av0} after making the following changes. Except for $\mathbb{P}(W_k\geq w)$, replace each instance of $W_j$, $v_j$, $H$, $\boldsymbol\phi^*$ and $c_0$ with $(\overline{W}_j,\overline{\tau}_j)$, $\widehat{v}_j$, $\overline{H}$, $\overline{\boldsymbol\phi}$ and $(c_0,0)$, respectively, for $j=0,i,i+1,k-2,k-1,k$. Replace each instance of $\mathbb{P}(W_k\geq w)$ with $\mathbb{P}(\overline{W}_k\geq w)$. The final result is
\begin{equation*}
\underset{\boldsymbol\pi_0,\boldsymbol\pi_1,...,\boldsymbol\pi_{k-1}}{\sup}\mathbb{P}(\overline{W}_k\geq w)=\widehat{v}_0(c_0,0)=\overline{v}_0(c_0).
\end{equation*}

\end{appendices}

\begin{algorithm}
\caption{Compute $\mathbb{P}(W_k\geq w)$ given $\boldsymbol\pi_i$ for $i=0,1,...,k-1$}
\label{alg:usim}
\begin{algorithmic}
\Require $n=2$, $N\in\mathbb{N}$ sufficiently large
\Require $X_{1i}$ are independent for $i=0,1,...,k-1$
\Require $X_{2i}=1+r$, $r>-1$ for $i=0,1,...,k-1$
\Require $\boldsymbol\pi_{i}=(q_i(W_i),1-q_i(W_i))$ for $i=0,1,...,k-1$
\State $m\gets 0$\Comment{initialize m}
\While{$m\leq N$}
\State $m\gets m+1$
\State $i\gets 0$ \Comment{initialize i}
\State $W\gets c_0$ \Comment{initialize $W$}
\While{$i\leq k$}
\State $i\gets i+1$
\State $X$ is a realization of $X_{1,i-1}$
\State $W\gets (q_{i-1}(W)X+(1-q_{i-1}(W))(1+r))W+c_i$ \Comment{computes $W_i$}
\EndWhile
\State $b_m\gets\begin{cases}1,&W\geq w\\0,&\text{otherwise}\end{cases}$
\EndWhile
\State $\mathbb{P}(W_k\geq w)\gets\frac{1}{N}\sum_{m=1}^Nb_m$\\
\Return{$\mathbb{P}(W_k\geq w)$}
\end{algorithmic}
\end{algorithm}

\begin{algorithm}
\caption{Compute $\mathbb{P}(\overline{W}_k\geq w)$ given $\boldsymbol\pi_i$ for $i=0,1,...,k-1$}
\label{alg:uhatsim}
\begin{algorithmic}
\Require $n=2$, $N\in\mathbb{N}$ sufficiently large
\Require $X_{1i}$ are independent for $i=0,1,...,k-1$
\Require $X_{2i}=1+r$, $r>-1$ for $i=0,1,...,k-1$
\Require $\boldsymbol\pi_{i}=(q_i(\overline{W}_i),1-q_i(\overline{W}_i))$ for $i=0,1,...,k-1$
\State $m\gets 0$\Comment{initialize m}
\While{$m\leq N$}
\State $m\gets m+1$
\State $i\gets 0$ \Comment{initialize i}
\State $\overline{W}\gets c_0$ \Comment{initialize $\overline{W}$}
\State $\widehat{\tau}$ is a realization of $\tau$
\While{$i\leq k$}
\State $i\gets i+1$
\State $X$ is a realization of $X_{1,i-1}$
\State $\overline{W}\gets (1-\mathcal{X}_{(0,i]}(\widehat{\tau}))[(q_{i-1}(\overline{W})X+(1-q_{i-1}(\overline{W}))(1+r))\overline{W}+c_i]$
\State $\quad\quad\quad+\ \mathcal{X}_{(0,i]}(\widehat{\tau})\overline{W}$ \Comment{computes $\overline{W}_i$}
\EndWhile
\State $b_m\gets\begin{cases}1,&\overline{W}\geq w\\0,&\text{otherwise}\end{cases}$
\EndWhile
\State $\mathbb{P}(\overline{W}_k\geq w)\gets\frac{1}{N}\sum_{m=1}^Nb_m$\\
\Return{$\mathbb{P}(\overline{W}_k\geq w)$}
\end{algorithmic}
\end{algorithm}

\begin{algorithm}
\caption{Compute $v_i$ and optimal $\boldsymbol\pi_{i}$ for $i=0,1,...,k-1$}
\label{alg:v}
\begin{algorithmic}
\Require $n=2$, $M\in\mathbb{N}$ sufficiently large
\Require $X_{1i}$ are iid and continuous with pdf $f$ and cdf $F$ for $i=0,1,...,k-1$
\Require $X_{2i}=1+r$, $r\geq0$ for $i=0,1,...,k-1$
\Require $G_1=\{.01,.02,...,.99\}$
\State $i\gets k-1$ \Comment{initialize i}
\State $D_i\gets\Big\{\frac{mw_i}{M}:m=1,...,2M\Big\}$\Comment{initialize $D_i$}
\State $v_i(x)\gets\begin{cases}1-F(w_{k-1}(1+r)/x)&x\in D_i\cap(0,w_{k-1})\\1&x\in D_i\cap[w_{k-1},\infty)\end{cases}$ \Comment{see \eqref{eq:vkm1}}
\While{$i>0$}
\State $i\gets i-1$
\State $D_i\gets\Big\{\frac{mw_i}{M}:m=1,...,2M\Big\}$
\For{$x\in D_i$}
\State $\theta\gets(1+r)x+c_{i+1}$ \Comment{temporary variable}
\State $y^*\gets\underset{y\in D_{i+1}\cap(0,\theta]}{\arg\min}\theta-y$
\State $q_i^*(x)\gets 0$ \Comment{initial proposal for $q_i^*(x)$}
\State $v_i(x)\gets \begin{cases}0&\theta<\underset{y\in D_{i+1}}{\min} y\\v_{i+1}(y^*)&\underset{y\in D_{i+1}}{\min}  y\leq\theta<w_{i+1}\\1&w_{i+1}\leq\theta\end{cases}$\Comment{initial proposal for $v_i(x)$}
\State $q_1\gets\underset{q\in G_1}{\arg\max}\ \eqref{eq:vqfTS}$
\State $G_2\gets\{q_1\pm .001m: m=0,1,...,9\}$
\State $q_2\gets\underset{q\in G_2}{\arg\max}\ \eqref{eq:vqfTS}$
\State $G_3\gets\{q_2+.0001m:m=-9,-8,...,10\}$
\State $q_3\gets\underset{q\in G_3}{\arg\max}\ \eqref{eq:vqfTS}$
\State $V\gets \eqref{eq:vqfTS}\vert_{q=q_3}$
\If{$v_{i}(x)<V$}
\State $q_i^*(x)\gets q_3$ 
\State $v_i(x)\gets V$
\EndIf
\EndFor
\EndWhile\\
\Return{$v_i(W_i),\ \boldsymbol\pi_i=(q_i^*(W_i),1-q_i^*(W_i))$ for $W_i\in D_i$ and $i=0,1,...,k-1$}
\end{algorithmic}
\end{algorithm}

\begin{algorithm}
\caption{Compute $\overline{v}_i$ and optimal $\boldsymbol\pi_{i}$ for $i=0,1,...,k-1$}
\label{alg:vhat}
\begin{algorithmic}
\Require $n=2$, $M\in\mathbb{N}$ sufficiently large
\Require $X_{1i}$ are iid and continuous with pdf $f$ and cdf $F$ for $i=0,1,...,k-1$
\Require $X_{2i}=1+r$, $r\geq0$ for $i=0,1,...,k-1$
\Require $G_1=\{.01,.02,...,.99\}$
\State $i\gets k-1$ \Comment{initialize i}
\State $D_i\gets\Big\{\frac{mw_i}{M}:m=1,...,2M\Big\}$\Comment{initialize $D_i$}
\State $\overline{v}_i(x)\gets\begin{cases}(1-p_{k-1})[1-F(w_{k-1}(1+r)/x)]\\+\ p_{k-1}v_k(x)&x\in D_i\cap(0,w_{k-1})\\1&x\in D_i\cap[w_{k-1},\infty)\end{cases}$ 
\While{$i>0$}
\State $i\gets i-1$
\State $D_i\gets\Big\{\frac{mw_i}{M}:m=1,...,2M\Big\}$
\For{$x\in D_i$}
\State $\theta\gets(1+r)x+c_{i+1}$ \Comment{temporary variable}
\State $y^*\gets\underset{y\in D_{i+1}\cap(0,\theta]}{\arg\min}\theta-y$
\State $q_i^*(x)\gets 0$ \Comment{initial proposal for $q_i^*(x)$}
\State $\overline{v}_i(x)\gets \begin{cases}p_iv_k(x)&\theta<\underset{y\in D_{i+1}}{\min} y\\(1-p_i)\overline{v}_{i+1}(y^*)+p_iv_k(x)&\underset{y\in D_{i+1}}{\min}  y\leq\theta<w_{i+1}\\1&w_{i+1}\leq\theta\end{cases}$
\State $q_1\gets\underset{q\in G_1}{\arg\max}\ \eqref{eq:vqfTS}\vert_{v=\overline{v}}$
\State $G_2\gets\{q_1\pm .001m: m=0,1,...,9\}$
\State $q_2\gets\underset{q\in G_2}{\arg\max}\ \eqref{eq:vqfTS}\vert_{v=\overline{v}}$
\State $G_3\gets\{q_2+.0001m:m=-9,-8,...,10\}$
\State $q_3\gets\underset{q\in G_3}{\arg\max}\ \eqref{eq:vqfTS}\vert_{v=\overline{v}}$
\State $V\gets (1-p_i)\eqref{eq:vqfTS}\vert_{v=\overline{v},q=q_3}+p_iv_k(x)$
\If{$\overline{v}_{i}(x)<V$}
\State $q_i^*(x)\gets q_3$ 
\State $\overline{v}_i(x)\gets V$
\EndIf
\EndFor
\EndWhile\\
\Return{$\overline{v}_i(\overline{W}_i),\ \boldsymbol\pi_i=(q_i^*(\overline{W}_i),1-q_i^*(\overline{W}_i))$ for $\overline{W}_i\in D_i$ and $i=0,1,...,k-1$}
\end{algorithmic}
\end{algorithm}

\clearpage
\bibliography{sn-bibliography}


\end{document}